\def\rfr#1{eq. (\ref{#1})}

\def\dert#1#2{\frac{{{d}}{#1}}{{{d}}{#2}}}
\def\virg#1{``#1''}

\def\eqi{\begin{equation}}
\def\eqf{\end{equation}}
\def\eqia{\begin{eqnarray}}
\def\eqfa{\end{eqnarray}}
\def\Om{\mathit{\Omega}}
\def\rp#1#2{{#1\over#2}} \def\lb#1{\label{#1}}

\def\bds#1{\boldsymbol{#1}}
\def\rg{R_{\rm r}}
\def\mg{m_{\rm r}}
\def\Ir{I_{\rm r}}

\def\Or{\Om_{\rm r}}

\def\DO{\Delta\Om}

\def\ee{e^2}

\def\ton#1{\left(#1\right)}
\def\qua#1{\left[#1\right]}
\def\grf#1{\left\{#1\right\}}
\def\ang#1{\left\langle #1\right\rangle}

\documentclass[11pt]{article}
\usepackage{url}
\usepackage{amsmath,amsthm,amscd,amssymb}
\usepackage{latexsym,w-greek,wasysym}
\usepackage{graphicx,epsfig}
\usepackage{hyperref}
\RequirePackage{color}

\begin{document}

\title{Orbital perturbations due to massive rings}

\author{L. Iorio \\ Ministero dell'Istruzione, dell'Universit\`{a} e della Ricerca (M.I.U.R.)-Istruzione \\ Fellow of the Royal Astronomical Society (F.R.A.S.) \\
 International Institute for Theoretical Physics and
Advanced Mathematics Einstein-Galilei \\ Permanent address: Viale Unit$\grave{\rm a}$ di Italia 68
70125 Bari (BA), Italy \\ email: lorenzo.iorio@libero.it}

\maketitle

\begin{abstract}
We analytically work out the long-term orbital perturbations induced by a homogeneous circular ring of radius $R_{\rm r}$ and mass $\mg$ on the motion of a test particle in the cases (I): $r > R_{\rm r}$ and (II): $r < R_{\rm r}$. In order to extend the validity of our analysis to the orbital configurations of, e.g., some proposed spacecraft-based mission for fundamental physics like LISA and ASTROD, of possible annuli around the supermassive black hole in Sgr A$^{\ast}$ coming from tidal disruptions of incoming gas clouds, and to the effect of artificial space debris belts around the Earth, we do not restrict ourselves to the case in which the ring and the orbit of the perturbed particle lie just in the same plane. From the corrections $\Delta\dot\varpi^{(\rm meas)}$ to the standard secular perihelion precessions, recently determined by a team of astronomers for some planets of the Solar System, we infer upper bounds on  $m_{\rm r}$  for various putative and known annular matter distributions of natural origin (close circumsolar ring with $R_{\rm r} = 0.02-0.13\ {\rm au}$, dust ring with $R_{\rm r} = 1\ {\rm au}$, minor asteroids, Trans-Neptunian Objects).
We find $m_{\rm r}\leq 1.4\times 10^{-4}\ m_{\oplus}$ (circumsolar ring with $R_{\rm r} = 0.02$ au), $m_{\rm r}\leq 2.6\times 10^{-6}\ m_{\oplus}$ (circumsolar ring with $R_{\rm r} = 0.13$ au), $m_{\rm r}\leq 8.8\times 10^{-7}\ m_{\oplus}$ (ring with $R_{\rm r} = 1$ au), $m_{\rm r}\leq 7.3\times 10^{-12}\ M_{\odot}$ (asteroidal ring with $R_{\rm r} = 2.80$ au), $m_{\rm r}\leq 1.1\times 10^{-11}\ M_{\odot}$ (asteroidal ring with $R_{\rm r} = 3.14$ au), $m_{\rm r}\leq 2.0\times 10^{-8}\ M_{\odot}$ (TNOs ring with $R_{\rm r} = 43$ au).
In principle, our analysis is valid both for baryonic and non-baryonic Dark Matter distributions.
\end{abstract}

%\keywords{gravitation-Oort Cloud-celestial mechanics}
%\keywords{black hole physics-Galaxy:center-relativity-techniques: radial velocities}
%\keywords{Experimental studies of gravity; Experimental tests of gravitational theories; Ephemerides, almanacs, and calendars}
%PACS: 04.80.-y; 04.80.Cc; 95.10.Km
%
%\keywords{gravitation - celestial mechanics - ephemerides - planet$-$disc interactions - Kuiper belt: general - minor planets, asteroids: general}
\centerline
{Keywords: Gravity - Asteroid Belt - Interplanetary Dust - Perturbation Methods}
\section{Introduction}
The giant  planets of the Solar System are surrounded by rings \cite{miner07}. They have widely been studied and have been the object of
numerous scientific spacecraft-based missions \cite{elliot77,johnson79,smith79,horn90,daubar99,porco04}.
As we will see in more details in Section \ref{review}, there is the possibility that circumsolar massive rings also exist at small distances from the Sun \cite{brecher79,rawal011}.
Moreover, the dynamical action of the ensemble of the minor asteroids of the asteroid belt between Mars and Jupiter can roughly be modeled as due to
a continuous ring \cite{krasinsky02}. The same holds for the belt \cite{edgeworth43,kuiper51,fernandez80} of Trans Neptunian Objects (TNOs) \cite{pitjeva010a,pitjeva010b}.

Recent observations suggest that similar structures exist in other stellar systems as well. \cite{marshall011} detected a ring-like structure   at about 70 au from the Sun-type star HD 207129 \cite{vanleeuwen07} with the IR observations collected by the instruments PACS
\cite{poglitsch010} and SPIRE \cite{griffin010} of the Herschel space observatory \cite{pilbratt010}. For studies on Kuiper belt-like structures at about $30-50$ au or beyond around other nearby Solar-type main-sequence stars, see, e.g., \cite{nilsson010} and \cite{shannon011}. In particular, \cite{nilsson010}
detected 10 exo-Kuiper belts with masses ranging from to a minimum\footnote{A lower bound of $76\ m_{\leftmoon}$ was inferred by \cite{nilsson010}  for the star HD 95086 \cite{dezeeuw99}.} of $0.3\ m_{\leftmoon}$ to a maximum of $48\ m_{\leftmoon}$  by means of far-IR observations from the APEX telescope \cite{gusten06}.

On the other hand, rings are common features in different astrophysical objects as well. \cite{nayakshin012} suggested that supermassive black holes located at the center of galaxies may be shrouded by super-Oort clouds of comets and asteroids. \cite{ansorg05}  performed  numerical solutions to the problem of black holes surrounded by uniformly rotating rings in axially symmetric, stationary spacetimes. \cite{karas04} reviewed the properties of gravitating discs around black holes.
Moreover, annular structures occur also in several ring galaxies (or R galaxies); they are objects with approximate
elliptical rings and no luminous matter visible in their interiors \cite{theys76,theys77}. Furthermore,
in some circumstances, as the result of interactions between galaxies, a ring of gas and stars is
formed and rotates over the poles of a galaxy, resulting in the polar-ring galaxies \cite{whitmore90}.

In addition to annular structures of natural origin, in the last decades analogous matter distributions of anthropogenic objects formed around the Earth \cite{anz95,flury95}. Indeed, as first pointed out by \cite{kessler78}, collisions among artificial satellites would produce orbiting fragments, each of which would increase the probability of further collisions, leading to the growth of a belt of debris around our planet in the Low Earth Orbit (LEO) region\footnote{Satellites moving in it have typically  altitudes  $h\lesssim 300-1,500$ km \cite[p. 2]{montenbruck00} .} after just decades. \cite{liou08} showed that the growth of such a debris population is primarily driven by high collision activities around $900-1,000$ km altitude. Radar-based observational campaigns were implemented to  characterize the orbital debris environment \cite{stansbery95}. Much more long-lived debris equatorial rings, with lifetimes of the order of millennia, may form  in the Geostationary Earth Orbit (GEO) region\footnote{Geostationary spacecraft orbit at $h=35,800$ km  \cite[p. 4]{montenbruck00}.} as well \cite{friesen92,debris95,anselmo02}; see, e.g. Figure 1.1 in \cite[p. 1]{montenbruck00}. For a recent overview, see \cite{klinkrad06}.

Self-gravitating toroidal fluid configurations (without a central
body) in Newtonian gravity were  analytically
studied by \cite{poincare85a,poincare85b,poincare85c}, \cite{dyson92,dyson93}, \cite{kowalewsky95}, and \cite{lichtenstein33}. In
particular, \cite{dyson92,dyson93} set up an approximation scheme of
uniformly rotating, homogeneous, and axisymmetric rings
which turned out to be an extremely good approximation for
thin rings; see \cite{ansorg03a}.
For various physical properties of flat and toroidal ring models, see \cite{letelier07} and \cite{vogt09}, who exploited the disk schemes by \cite{morgan69}, \cite{kuzmin56} and \cite{toomre63}. For other approaches, see  \cite{ciotti07} and \cite{ciotti08}.
\cite{petroff08} analytically considered the problem of a uniformly rotating, self-gravitating ring without a central body in Newtonian gravity.
A general relativistic toroid around a black hole was studied by \cite{nishida94}.
\cite{vogt05} and \cite{ujevic011} considered three-dimensional models for the gravitational field of rings in the context of general relativity.
Uniformly rotating, homogeneous and axisymmetric relativistic fluid bodies with a toroidal shape were investigated by \cite{ansorg03b}.
\cite{horatschek010} considered uniformly rotating homogeneous rings in post-Newtonian gravity. Uniformly rotating rings in general relativity were investigated also by \cite{fischer05}.
\cite{kleinwaechter011} studied the black hole limit of rotating discs and rings.

As far as the orbital perturbations induced by such mass configurations, \cite[p. 58]{kellog29} and  \cite[p. 195]{macmillan30} treated the problem of the attraction of a
continuous circular massive ring, and showed
that its potential can be approximated by a complete elliptic integral of the first kind \cite{abramowitz72}. Actually,
Gauss had already solved the problem and he used it to state his famous averaging theorem: replacing
a perturbing body by an equivalent continuous mass spread over its orbit does not change the secular
effects while it removes the periodic terms of the perturbation \cite{broucke05}. See also \cite{Eck02}. \cite{Zyp06} analytically treated the equivalent problem of calculating the off-axis electric field of a ring of charge. Numerical computation of the acceleration due to a uniform ring can be found in \cite{Fuku010}.
The perturbing acceleration on an inner particle due to various external axisymmetric mass distributions (e.g. different kinds of thin rings, uniform and non-uniform disk, torus) was worked out by \cite{anderson02}, \cite{nieto05}, \cite{bertolami06}, \cite{dediego06} in the framework of the Pioneer anomaly \cite{anderson02}. \cite{Vash012} analytically worked out the force function of a slightly elliptical Gaussian ring,
and treated its generalization to a nearly coplanar system of rings as well.
For some approximate calculations of the orbital perturbations on the elliptical motion of a test particle moving inside annular distributions, see, e.g., \cite{iorio07} and \cite{kuchynka010}.
The nature of test particle motion in the presence of a rotating ring of self-gravitating matter around a Kerr black hole was investigated by \cite{khanna92}.
The perturbations by a distribution of stars in the Galactic Center were investigated by \cite{sadeghian011}.

In this paper, we  analytically work out the long-term, i.e. averaged over one full orbital period, perturbations of a test particle moving about a central body and acted upon by a circular massive ring. More specifically, after having reviewed various baryonic and non-baryonic scenarios involving annular matter distributions in Section \ref{review}, we  consider the potential induced by the material annulus in the general case of an arbitrary ring-orbit configuration (Section \ref{potenza}). Then, we analytically compute the long-term variations of all the osculating Keplerian orbital elements of a test particle due to a circular ring of homogenous linear density. As a by-product, we preliminarily put model-independent, dynamical constraints on the mass of several annular matter distributions from the orbital motions of some of the planets of the Solar System. In particular, we use the latest determinations of the supplementary secular precessions $\Delta\dot\varpi^{(\rm meas)}$ of the planetary perihelia \cite{fienga011}.  In applying our results to the Solar System, we  assume that the perturbed orbital motions occurs in the ring's plane itself. Table \ref{tavola} shows that it can be considered a rather reasonable approximation for those planets of the Solar System for which \cite{fienga011} determined $\Delta\dot\varpi^{(\rm meas)}$ by using the mean Earth's equator at the epoch J$2000.0$ as reference $\{x,y\}$ plane.
\begin{table*}[ht!]
\caption{Inclination $I$, longitude of the ascending node $\Om$  and argument of pericenter $\omega$  of the planets of the Solar System for which \cite{fienga011} determined the supplementary perihelion  precessions $\Delta\dot\varpi^{(\rm meas)}$ with respect to the mean Earth's equator and equinox at J$2000.0$. $I$ and $\Om$ determine the orientation of the orbital plane in space, while $\omega$ yields the orientation of the orbit in its plane \cite{murray99}.
}\label{tavola}
\centering
\bigskip
\begin{tabular}{lllllll}
\hline\noalign{\smallskip}
 & Mercury & Venus & Earth & Mars & Jupiter  & Saturn\\
\noalign{\smallskip}\hline\noalign{\smallskip}
$I$ (deg) & $28.6$ & $24.4$ & $23.4$ &  $24.7$ & $23.2$ & $22.5$\\
$\Om$ (deg) & $10.9$ & $7.9$ & $0.0$  & $3.4$ & $3.2$ & $5.9$\\
$\omega$ (deg) & $-295.7$ & $145.2$ & $76.7$  & $-27.5$ & $10.4$ & $87.1$\\
\noalign{\smallskip}\hline\noalign{\smallskip}
\end{tabular}
\end{table*}
We consider both the cases in which the planet moves outside the ring (Section \ref{interno}), and inside it (Section \ref{esterno}).
As far as the hypothetical annuli are concerned, a complementary approach which may be followed consists of explicitly modeling them and solving for a dedicated parameter in fitting the models to the observations.

We remark that our analysis is not necessarily limited to the major natural bodies of the Solar System (and of other stellar systems). Indeed, annular massive distributions like, e.g., the minor asteroid ensemble  have an impact on several proposed space-based missions for fundamental physics like ASTROD \cite{ni08,men010,dong011}, LISA \cite{faller84,faller85,povoleri06,gath010}, GAIA \cite{turon05,lindegren010} relying upon spacecraft in Solar orbits of about 1 au. Moreover, man-made space debris belts orbiting the Earth may  affect the orbital motion of terrestrial artificial satellites of interest for fundamental physics as well.
Section \ref{conclusioni} summarizes our findings.
%In the Appendix we treat the general case of an arbitrary ring-orbit configuration, particularly suited when anthropogenic test particles perturbed by natural %and artificial rings are involved.
%
%

As a final remark, we point out that the present work has  no pretensions of exhaustiveness  about the whole ring problem.  In particular, it purposely does not deal with the domain of periodic orbits, where the potential in closed form expressions is used. For such a potential, there are orbits rather complicated, which likely could not be represented with only a few terms of the Legendre expansion used here. These orbits have a high variable orbital elements, and an averaging over the true anomaly like that performed in this paper gives no actual information.
For studies on some general features of the dynamics around a massive annulus, like  orbit stability for various configurations and the equilibrium of the system, see. e.g., \cite{broucke05,Albe07,Aze07a,Aze07b,ramos011, tresaco011} and \cite{tresaco012}.

Moreover, we do not pursue elegant and compact  formulations which, if on the one hand may be of interest for the specialized reader, on the other hand may have little practical uses for the broad class of problems outlined here and for the allegedly broader audience interested in them.
\section{Baryonic and non-baryonic matter distributions in the Solar System}\lb{review}
\subsection{Baryonic matter rings}
Is the Sun surrounded by a ring of rocky material lying at no more than a few Solar radii ($1\ R_{\odot}=0.00465\ {\rm au}$)?

Such a hypothesis, along with that of a putative intra-Mercurial planet named Vulcan, was put forth for the first time by \cite{leverrier59} in the form of a ring of asteroids,  known also as \virg{Vulcanoids} \cite{sandage00}, between Mercury and the Sun to explain the anomalous perihelion precession of Mercury of 38 degrees of arc per century (arcsec cty$^{-1}$) discovered by him in 1859. Later, \cite{newcomb82}, by including more observations in a new analysis of the motion of Mercury, came to an anomalous precession of\footnote{Its currently accepted value is $42.98$ arcsec cty$^{-1}$; see \cite{nobili86} for a discussion.} 43 arsec cty$^{-1}$. It was explained by \cite{seeliger06} in terms of a zodiacal matter  distribution whose gravitational attraction was equal to that of two ellipsoids, one inside Mercury's orbit and the other outside that of Venus. The hypothesis of a matter ring was supported also by \cite{poincare53} in a series of lectures given in 1906-1907 on the limits of the Newton's law of gravitation. As it is well known, \cite{einstein15} finally explained the anomalous perihelion precession of Mercury within the framework of its general theory of relativity which does not contain any ad-hoc, adjustable parameter.
The possible existence of a circumsolar belt of small asteroids at about $0.1$ au was revamped by \cite{courten72} as a consequence of a reanalysis of some photographic plates taken during an eclipse in 1970. For more details, see \cite{seargent011}.
\cite{evans99} pointed out that  Vulcanoids not smaller than 100 m may exist in a gravitationally stable band inside the orbit of Mercury, at distances of $0.06-0.21$ au from the Sun, with an inclination to the ecliptic not exceeding 10 deg. See also \cite{lebofsky75,vokrouhlicky99,durda00}. Their existence would be in agreement with the predictions of the scale relativity theory \cite{nottale97}. \cite{schumacher01}, in their unsuccessful attempt to directly detect Vulcanoids in coronagraph images from the instrument LASCO \cite{brueckner95} carried onboard the spacecraft SoHO \cite{domingo95}, placed an upper bound on the size of such hypothetical objects which should not exceed 60 km. In a previous SoHO/LASCO-based investigation, \cite{durda00} concluded that the present-day population of Vulcanoids larger than 1 km  should amount to less than about $1,800-42,000$ objects.

The discovery of rings around some of the outer giant planets of the Solar System in the end of 70s \cite{elliot77,johnson79,smith79} enforced the possibility that a circumsolar massive ring exists as well \cite{brecher79}. Various evolutionary, physical, chemical and observational considerations have been provided about such a hypothesis. According to \cite{brecher79}, the constituents would be $\gtrsim 10$ km size boulders made of refractory material like graphite. As far as the radius $R_{\rm r}$ of a permanent ring structure around the Sun is concerned, the values \cite{brecher79}
\eqi
\begin{array}{lll}
R_{\rm r} & = & 0.02\ {\rm au}= 4.3\ R_{\odot}, \\ \\
R_{\rm r} & = & 0.13\ {\rm au} =27.9 \ R_{\odot}, \\ \\
\end{array}
\eqf
were proposed. The total mass would be much less than  $m_{\rm r} \lesssim 6\times 10^{22}$ kg \cite{brecher79}. \cite{belton66,belton67}  theoretically predicted the existence of
a circumsolar dust ring close to the Sun; for a review on the formation and possible observation of a dust ring, see \cite{kimura98}. Direct searches for a hypothetical circumsolar ring by exploiting its electromagnetic emission in the infrared (IR) portion of the spectrum
were conducted yielding contradictory results \cite{singh04}, especially as far as the nature and the dimensions of the putative constituents of the ring are concerned.  Earlier  observations of the F-corona brightness
enhancement near $4\ R_{\odot}$  by \cite{macqueen68} and
\cite{peterson67,peterson69} supported the ring  hypothesis. \cite{rao81} attempted to detect the existence of such a ring structure utilizing the IR observations taken during optimal viewing conditions of the total Solar eclipse of 16 February 1980. The near-IR observations by \cite{mizutani84},  who exploited the 1983 Solar eclipse, supported the existence of a circumsolar ring of dust lying approximately in the ecliptic plane at about $4\ R_{\odot}$. \cite{hodapp92}, in their  search for excess IR emission in the Solar equatorial plane during the eclipse of 11 July 1991, concluded that the circumsolar dust ring may be a transient feature, perhaps due to the injection of dust into near-Solar space by a sun-grazing comet. \cite{lamy92} argued against the existence of a circumsolar dust ring. According to \cite{debi95}, the ring's constituents must be fine particles, instead of 10 km sized boulders at
$4\ R_{\odot}$ from the Sun. \cite{mann05} concluded that it is not reasonable to suggest the presence of Si nanoparticles in the vicinity of the Sun. The hypothesis of a massive circumsolar ring made of macroscopic chunks of matter was recently reanalyzed by \cite{rawal011} who showed that, while a possible Solar ring structure at  $4\ R_{\odot}$ would not be stable over the age of the Solar System because of the Alfven drag \cite{drell65} produced during even moderate Solar magnetic storms, a ring at $\sim 27\ R_{\odot}$ would, instead, be stable even during very intense Solar magnetic activity.

\cite{leinert07} found evidence  for a dust ring associated with the orbit of Venus in the data from the Helios-2 probe \cite{desai81}.	
\cite{jackson89} and \cite{dermott94} pointed out the possible existence of a circumsolar ring of asteroidal dust with $R_{\rm r}\sim 1\ {\rm au}$; the azimuthal structure of this  ring was predicted to be asymmetric, with the region trailing the Earth being substantially more dense than that in the leading direction. \cite{reach94}, who analyzed data collected by the instrument Diffuse Infrared Background Experiment (DIRBE) \cite{hauser93} carried onboard the  COsmic Background Explorer (COBE) satellite \cite{cobe}, confirmed the existence of such a ring.  \cite{reach010} further investigated the structure of the Earth's circumsolar dust ring.
For a review about dust bands in the Solar System and in extra-Solar planetary systems, see \cite{mann06}.

Other matter distributions, residing far from the Sun, which can be treated with the approach outlined here are the ring of  minor asteroids between Mars and Jupiter \cite{krasinsky02}, and the TNOs' belt  \cite{edgeworth43,kuiper51,fernandez80}.
The action of the minor asteroids was modeled as due to a static circular ring at 2.8 au \cite{krasinsky02,standish03,konopliv06,fienga08,kuchynka010} or at 3.14 au \cite{pitjeva05a,fienga09,kuchynka09,pitjeva010b}.
\cite{pitjeva010a,pitjeva010b} modeled the TNOs belt as a circular massive ring with $R_{\rm r}=43\ {\rm au}$.
\subsection{Non-baryonic Dark Matter distributions}
Until now we only considered rings made of ordinary baryonic  matter. In principle, our analysis can  extend to
non-baryonic Dark Matter (DM) as well.

Its existence was postulated long ago to explain the discrepancy between the observed
kinematics of some components of astrophysical systems like clusters of galaxies \cite{zwicky33} and spiral
galaxies \cite{bosma81,rubin82}, and the predicted one on the basis of the Newtonian dynamics and the matter
directly detected from the emitted electromagnetic radiation (visible stars and gas clouds).
\cite{zwicky37} postulated the existence of undetected, baryonic matter; today \cite{rubin83} it is believed
that the hidden mass is constituted by non-baryonic, weakly interacting particles in order to
cope with certain issues pertaining galaxy and galaxy clusters formation \cite{chiu03}, the Cosmic Microwave Background (CMB) \cite{komatsu09} and primordial
nucleosynthesis \cite{tytler00}. On cosmological scales, DM accounts for about $23\%$ of the mass-energy
content of the Universe \cite{hinshaw09}. A widely popular generic class of new-particle candidates is the
Weakly Interacting Massive Particle (WIMP) scenario \cite{steigman85,jungman96}. In it, DM
annihilates with itself and interacts with the rest of the Standard Model (SM) via the weak
interaction. The WIMP is typically defined as a stable, electrically neutral, massive particle
which arises naturally in supersymmetric SM extensions \cite{haber85}. A pair of WIMPs can annihilate,
producing ordinary particles and gamma rays.
\subsubsection{WIMPs}
DM particles in the Galactic halo, traversed by the Solar System during its Galactic journey, can end up on orbits gravitationally bound to the Solar System via purely gravitational three-body interactions with the planets \cite{gould91}. However, when only gravitational interaction is considered,  the inverse process of capture from the Galactic halo, leading to ejection of dark-matter particles from the Solar System, must be taken into account as well \cite{gould91,lundberg04,peter09a}. As a result, a detailed balance holds, so that a there is a rather small limiting density $\rho_{\rm DM}$ of dark matter particles in the Solar System, not particularly larger than the mean Galactic value $\rho_{\rm DM}\sim 4\times 10^{-25}\ {\rm g\ cm^{-3}}$ \cite{bertone05}. There are, however, different points of view on that issue according to which the situation with the true value for typical life-time of the captured DM in the Solar System would be far from being clear; see \cite{khriplovich09,khriplovich010,khriplovich011}.
  An increase in the population of DM particles bound to the Solar System can occur only if physical mechanisms other than pure gravitational scattering are at work like e.g., weak interactions with atoms in the Sun or the planets \cite{edsjo010}.
  Within such a framework, \cite{damour98} investigated a scenario in which WIMPs undergo weak scattering with atomic nuclei \cite{gould88} in  layers within the Sun close to its surface in such a way that they lose just enough energy to travel along closed Earth-crossing orbits with high eccentricities. Following such trajectories crossing the Solar surface, most of WIMPs are doomed to experience a second scattering event after just $10^3-10^4$ revolutions which would significantly reduce their semimajor axes $a_{\rm W}$: they will end up in the Sun's core where they will ultimately annihilate with each other. However, if the layer of Solar matter traversed is quite thin, the lifetime of the grazing WIMPs gets larger. Moreover, during such an increased time the WIMPs experience small orbital perturbations by other planets of the Solar System which, through the \cite{kozai62} mechanism, stabilize them in stable orbital configurations at 1 au which do not intersect the Sun anymore over times comparable to the Solar System's lifetime \cite{damour99}. More precisely, \cite{damour99} obtained that the long-term survival of such WIMPs orbits is greater than 4.5 Gyr if their semimajor axes are smaller than half the semimajor axis of Jupiter, i.e. for  $a_{\rm W}< a_{\jupiter}/2=2.6\ {\rm au}$. \cite{peter09b} further studied such a scenario: anyway, according to  \cite{peter09b}, the overall effect on the bound DM density should be small.
  The presence of such a stable WIMPs distribution at the Earth's location would have a great impact for the laboratory-based experimental searches of DM like, e.g., CDMSI \cite{akerib03}, CDMSII \cite{ahmed010}, DAMA/NaI \cite{bernabei03} and its successor
DAMA/LIBRA \cite{bernabei08}, XENON10 \cite{angle09} and ZEPLIN III \cite{summer05}; for a recent introduction to dedicated DM experiments, see \cite{schnee011}. Thus, it is of great importance to put model-independent, dynamical constraints on the total mass of such a putative distribution of DM at 1 au.
\subsubsection{Mirror matter}
Other possible stable matter distributions made of another kind of DM, the mirror matter \cite{lee56,kobzarev66,pavsic74,blinnikov83,foot91,khlopov91,khlopov99,silagadze01,foot04,foot06,okun07,ciarcelluti010}, have been hypothesized. It arises if instead of (or in addition to) assuming a symmetry between bosons and fermions, i.e. supersymmetry, one assumes
that nature is parity symmetric.  For a popular overview of such a scenario, see \cite{foot02}. Mirror matter was supposed to exist in the Solar System; see \cite{ignatiev00,foot01a,foot01,silagadze02,silagadze05,foot03a,foot03b}.
\subsubsection{CUDOs}
Recently, a new class of exotic objects in the Solar System has been proposed: the COmpact Ultra Dense Objects (CUDOs) \cite{dietl011}.
They would be gravitationally self-bound objects made of non-baryonic
(dark) matter fermions of mass 250 GeV to 100 TeV, or from massless fermions hidden by vacuum structure of similar energy scale \cite{dietl011}. The density at the center of a typical CUDO is more than $10^{12}$ times higher than nuclear density, so that such extremely compact objects are small both in mass (sub-planetary, $M_{\rm C} < 0.1\ M_{\oplus}$) and radius ($R_{\rm C} < 10$ cm), making them difficult to observe by conventional astronomical methods \cite{dietl011}.
Impact signatures in Solar System bodies have recently been proposed to constrain CUDOs' local abundances \cite{labun011}.
There may exist also CUDOs of other types, such as supersymmetric Q-balls \cite{coleman85,kusenko97,kusenko98} or CUDOs composed of standard
model particles \cite{zhitnitsky06,cumberbatch08}. Q-ball collisions with Earth have been
considered by \cite{kusenko09}.
\section{Calculation of the potential of the ring for a generic orientation of its plane  in space and with respect to the orbit of the test particle}\lb{potenza}
For the sake of generality, in the following we will consider an arbitrary configuration of the ring-particle system. Indeed,  in several astronomical and astrophysical scenarios involving, e.g., binary pulsars, exoplanets, etc. the coordinate system\footnote{It is customarily adopted a system with the reference $z$ axis oriented along the line-of-sight,
from the object to the observer, and the reference $\{x,y\}$ plane coinciding with the usual plane of the sky which is tangential
to the celestial sphere at the position of the object. In it, the reference $x$ axis is directed towards the  North Celestial Pole (see, e.g., \cite[p. 287]{will93} and \cite[p. 446]{roy05}).} adopted  does not generally use any characteristic physical and/or orbital plane of the bodies of interest. Moreover, it is reasonable to expect that in several practical situations it cannot  be assumed that the perturbing ring structures share the same plane of the perturbed bodies investigated. A striking example of non-coplanarity of several test particles orbiting their common primary is given by the stellar system orbiting the supermassive black hole in Sgr A$^{\ast}$ at the center of the Milky Way \cite{gillessen09}; see \cite{sadeghian011} for a calculation of the orbital perturbations experienced by a S-type star due to a distribution of bodies. Recently, \cite{gillessen012} discovered a dense gas cloud with $m_{\rm c}=3m_{\oplus}$ which is rapidly approaching the accretion zone of Sgr A$^{\ast}$ on a highly elliptic orbit ($e_{\rm c}=0.94$) with an orbital period of 137 yr: it should reach the perinigricon in 2013. Over the past three years the cloud has begun to disrupt, likely by the tidal forces of the black hole \cite{gillessen012}; it might, thus, create an annular matter distribution perturbing the S-type stars orbiting Sgr A$^{\ast}$. Other examples are given by the spacecrafts of the LISA \cite{povoleri06} and ASTROD \cite{ni08} constellations, whose orbital planes are planned to have quite different nodes $\Om$ (see, e.g., \cite{li08}). In the case of the space debris belts surrounding the Earth \cite{klinkrad06}, the GEO ring lies in the equatorial plane \cite{friesen92}, contrary to the orbits of several satellites potentially of interest for fundamental physics and/or other scopes which supposedly are perturbed by it.

Let us choose a generic inertial frame endowed with a Cartesian coordinate system having the central body of mass $M$ located at its origin $O$. Let $\bds r$ be the position vector of the perturbed test particle, while $\bds R$ denotes the position vector of a  mass element $dm_{\rm r}$ of the annulus. In general, the annulus is inclined by an angle $I_{\rm r}$ to the reference $\{x,y\}$ plane, and its intersection with the latter one is displaced by an angle $\Om_{\rm r}$ with respect to the reference $x$ axis. Thus, $I_{\rm r}$ and $\Om_{\rm r}$ are the same for all the mass elements $dm_{\rm r}$ of the annulus. Their position along it is picked up by an angle $\varphi_{\rm r}$, measured in the annulus's plane  from its intersection with the reference $\{x,y\}$ plane. Thus, each $dm_{\rm r}$ has its own $\varphi_{\rm r}$. By assuming a homogeneous  matter distribution, it will be $dm_{\rm r} = \rho_{\rm r}R d\varphi_{\rm r}$, where $\rho_{\rm r}$ is the constant linear mass density.

Quite generally, the Cartesian components of $\bds r$ can straightforwardly be expressed in the usual way  \cite[p. 51]{murray99}
%
%
%
%
%\begin{equation}
%{\begin{array}{lll}
% x &=& r\left[\cos\Om\cos\left(\omega + f\right)\ -\cos I\sin\Om\sin\left(\omega + f\right)\right],\\  \\
  %
% y &=& r\left[\sin\Om\cos\left(\omega + f\right) + \cos I\cos\Om\sin\left(\omega + f\right)\right],\\  \\
 %
% z &=& r\sin I\sin\left(\omega + f\right),
%
%\end{array}}\lb{xyz}
% \end{equation}
 %
 %
 %
 %
  in terms of the longitude of the ascending node $\Om$,
  %\cite[p. 49]{murray99}
  the inclination $I$ to the reference $\{x,y\}$ plane,
  %\cite[p. 49]{murray99}
  the argument of pericenter $\omega$,
  %\cite[p. 49]{murray99}
  the true anomaly $f$,
  %\cite[p. 28]{murray99}
  which is the time-dependent, fast variable giving the instantaneous position of the test particle,
  %
  %
  %
  %along its Keplerian ellipse,
  %
  %
  %whose equation is \cite[p. 28]{murray99}
 %
 %
 %\eqi r =\rp{a(1-e^2)}{1 + e \cos f},\lb{kepla}\eqf
 %
 %
 %where
 the semimajor axis $a$, and the eccentricity $e$ of the Keplerian ellipse of the test particle.
 % \cite[p. 26]{murray99}

 Likewise, by assuming that the matter annulus has, in general, an elliptic shape, for the components of $\bds R$ we can write
\begin{equation}
{\begin{array}{lll}
 X &=& R\left(\cos\Om_{\rm r}\cos \varphi_{\rm r}\ -\cos I_{\rm r}\sin\Om_{\rm r}\sin \varphi_{\rm r}\right),\\  \\
 Y &=& R\left(\sin\Om_{\rm r}\cos \varphi_{\rm r} + \cos I_{\rm r}\cos\Om_{\rm r}\sin \varphi_{\rm r}\right),\\  \\
 Z &=& R\sin I_{\rm r}\sin \varphi_{\rm r}.
\end{array}}\lb{XYZ}
 \end{equation}
 Note that, in general, $R$ can vary along the massive annulus according to a relation analogous to the standard expression of the Keplerian ellipse. Here we will assume $R=
 R_{\rm r}$ throughout the ring.

 The potential exerted by a mass element $dm_{\rm r}$ in a point located at $\bds r$ is
 \eqi dU_{\rm r} = -\rp{Gdm_{\rm r}}{\left|\bds r-\bds R\right|} = - \rp{Gm_{\rm r}}{2\pi q}\rp{d\varphi_{\rm r}}{\sqrt{1+\alpha^2 -2\alpha\eta\left(\varphi_{\rm r}\right) }},\lb{potenziale}\eqf
 where
 \eqi
 \begin{array}{lll}
 ({\rm I})\ q & = & r,\ \alpha\doteq \rp{R_{\rm r}}{r}\ {\rm for}\ r > R_{\rm r}, \\ \\
 ({\rm II})\ q & = & R_{\rm r},\ \alpha\doteq \rp{r}{R_{\rm r}} \ {\rm for} \ r < R_{\rm r},
 \end{array}\lb{casi}
 \eqf
 and
 \eqi \eta\left(\varphi_{\rm r}\right)\doteq \bds{\hat{r}}\bds\cdot\bds{\hat{R}}\lb{coseno}\eqf
is the cosine of the angle between $\bds r$ and $\bds R$. It explicitly depends on $\varphi_{\rm r}$.
%, as it turns out by  computing it by means of \rfr{xyz} and \rfr{XYZ}.
 It can be written as
 \eqi \eta\left(\varphi_{\rm r}\right)={\mathcal{C}}\cos\varphi_{\rm r} + {\mathcal{S}}\sin\varphi_{\rm r},\eqf
with
\eqi
\begin{array}{lll}
\mathcal{C} & = & \cos\Delta\Om\cos u - \cos I\sin\Delta\Om\sin u, \\ \\
\mathcal{S}& = & \cos I\sin\Delta\Om\cos u + \ton{\cos I\cos I_{\rm r}\cos\Delta\Om+\sin I\sin I_{\rm r}}\sin u,
\end{array}\lb{coseno2}
\eqf
with $u\doteq \omega + f$.
%
%\eqi
%\begin{array}{lll}
%\eta\left(\varphi_{\rm r}\right) & = & \cos\varphi_{\rm r} \cos\left(\omega + f\right) \cos\Delta\Om +
%  \cos I \cos I_{\rm r} \cos\Delta\Om \sin\varphi_{\rm r} \sin\left(\omega + f\right) +\\ \\
  %
%   & + &\sin I \sin I_{\rm r} \sin\varphi_{\rm r} \sin\left(\omega + f\right) + \left[\cos I_{\rm r} \cos\left(\omega + f\right) \sin\varphi_{\rm r} -\right.\\ \\
  %
%  & - &\left.\cos I \cos\varphi_{\rm r} \sin\left(\omega + f\right)\right] \sin\Delta\Om.
%\end{array}\lb{coseno2}
%\eqf
%
%
%
From the point of view of the computation of the potential of the ring, $I,\Om,\omega,f,I_{\rm r},\Om_{\rm r}$ entering \rfr{coseno2} are to be considered as fixed parameters.

By expanding \rfr{potenziale} in powers of $\alpha<1$, it can be obtained
\eqi dU_{\rm r} = \left(\Psi_0 + \Psi_1 + \Psi_2 + \Psi_3 + \Psi_4 + \cdots\right)d\varphi_{\rm r},\lb{kalimba}\eqf
 with
 \eqi
 \begin{array}{lll}
 \Psi_ 0 & = & -\rp{Gm_{\rm r}}{2\pi q}, \\ \\
  \Psi_ 1 & = & -\rp{Gm_{\rm r}\eta}{2\pi q}\alpha, \\ \\
  \Psi_ 2 & = & \rp{Gm_{\rm r}}{4\pi q}\left(1 - 3\eta^2\right)\alpha^2, \\ \\
  \Psi_ 3 & = & \rp{3Gm_{\rm r}\eta}{4\pi q}\left(1 - \rp{5}{3}\eta^2\right)\alpha^3, \\ \\
  \Psi_ 4 & = & -\rp{3Gm_{\rm r}}{16\pi q}\left[1- 10 \eta^2\left(1 + \rp{35}{30}\eta^2\right) \right]\alpha^4.
 \end{array}\lb{super}
 \eqf
 The contribution of each term of \rfr{super} can be treated separately. First, the overall potential has to be computed by integrating over $\varphi_{\rm r}$
 \eqi U_j = \int_0^{2\pi}\Psi_j \left(\varphi_{\rm r}\right)d\varphi_{\rm r},\ j=0,1,2,\ldots\eqf
We have
\eqi
\begin{array}{lll}
U_0 & = & -\rp{Gm_{\rm r}}{q}, \\ \\
U_1 & = & 0, \\ \\
U_2 & = & \rp{Gm_{\rm r}}{2q}\left[1 - \rp{3}{2}\left(\mathcal{C}^2 + \mathcal{S}^2\right)\right]\alpha^2, \\ \\
U_3 & = & 0, \\ \\
U_4 & = & -\rp{3Gm_{\rm r}}{8 q}\left\{1 + \rp{35}{8}\left(\mathcal{C}^4 + \mathcal{S}^4\right) -5\left[ \mathcal{S}^2 + {\mathcal{C}}^2\left(1 - \rp{7}{4} \mathcal{S}^2\right) \right]\right\}\alpha^4.
\end{array}
\eqf

Then, the perturbations of the Keplerian orbital elements of the test particle induced by $U_j\left(t\right),\ j=0,1,2,\cdots$
can straightforwardly be worked out with, e.g., the Lagrange planetary equations \cite[pp. 251-252]{murray99}. To this aim, the averages over one full orbital period $P_{\rm b}$ must be worked out
as
\eqi\left\langle U_j\right\rangle = \left(\rp{1}{P_{\rm b}}\right)\int_0^{P_{\rm b}}U_j\left(t\right) dt,\ j=0,1,2,\ldots\lb{avera}, \eqf
where $U_j,\ j=0,1,2,\ldots$ are to be evaluated onto the unperturbed Keplerian ellipse of the moving test particle; as such, they are to be thought as functions of time $t$.
Using the eccentric anomaly $E$ \cite[p. 32]{murray99} as fast variable of integration in \rfr{avera} turns out to be computationally more convenient when $q=R_{\rm r}$ since $U_j\propto r^j,\ j=1,2,\ldots$.
%
%
%
%
%Useful conversion formulas are \cite[p. 33]{murray99}
%
%
%\eqi
%\begin{array}{lll}
%\cos f & = & \rp{\cos E - e}{1 - e \cos E}, \\ \\
%
%\sin f & = & \rp{\sqrt{1-e^2}\sin E}{1 - e \cos E}, \\ \\
%
%r & = & a\left(1 - e \cos E\right), \\ \\
%
%dt & = & \rp{\left(1 - e \cos E\right)}{n_{\rm b}}dE,
%\end{array}\lb{solite}
%\eqf
%
%
%where  $n_{\rm b} \doteq \sqrt{GM/a^3}$ is the Keplerian mean motion of the test particle related to its orbital period by $P_{\rm b}=2\pi/n_{\rm b}$ %\cite[pp. 29-30]{murray99}.
%
%
%
%

Instead, when $q=r$ and $U_j\propto r^{-(j+1)},\ j=0,1,2,\ldots$, the use of the true anomaly $f$
%
%
%
%
%, for which it is
%\cite[p. 41]{murray99}
%\eqi dt = \rp{(1-e^2)^{3/2}}{n_{\rm b}\left(1 + e\cos f\right)^2}df,\eqf
%
%
%
%
is computationally more suitable.

Performing the integrations of \rfr{avera} is, in general, quite cumbersome even for small values of $j$ in \rfr{super}.

If the annulus and the test particle's orbit are coplanar, the calculations are simpler. The potential of the ring can be worked out
by using the standard mathematical result, valid for $\alpha <1$,
\begin{align}
\rp{1}{2\pi}\int_0^{2\pi}\rp{d\xi}{\sqrt{1  + \alpha^2- 2\alpha\cos\xi}} \nonumber & =
\rp{1}{\pi}\left\{\rp{K\left[-\rp{4\alpha}{\left(1-\alpha\right)^2}\right]}{1-\alpha}+\rp{K\left[\rp{4\alpha}{\left(1+\alpha\right)^2}\right]}{1+\alpha}\right\}=\\ \nonumber \\
& = 1+\rp{1}{4}\alpha^2 + \rp{9}{64}\alpha^4 + \rp{25}{256}\alpha^6+\cdots,
\end{align}
where $K(k)$ is the complete elliptic integral of the first kind \cite{abramowitz72}.
Thus, the potential is
\eqi U_{\rm r}  =  -\rp{G m_{\rm r}}{2\pi q}\int_0^{2\pi}\rp{d\varphi_{\rm r}}{\sqrt{1 + \alpha^2 -2\alpha\cos\varphi_{\rm r}}}= -\rp{Gm_{\rm r}}{q}\left( 1 + \rp{1}{4}\alpha^2 + \rp{9}{64}\alpha^4 +\cdots\right).\lb{piatto}\eqf
\cite{lass83} yielded and expression in closed form, in terms of complete elliptic integrals. See also \cite{roth72,kasatkin05}, and \cite{Lask010} who treated the equivalent problem of the computation of the averaged
Hamiltonian in the case of a elliptical orbit perturbed by a circular
perturber.
It can be shown that \rfr{piatto} yields just the acceleration at $r$ obtained by \cite{fienga08}.
\section{The inner ring: $r > R_{\rm r}$ }\lb{interno}
In the case $({\rm I})$ of \rfr{casi}, it is
\eqi\left\langle U_0\right\rangle = -\rp{Gm_{\rm r}}{a}.\eqf
Thus,  \textcolor{black}{provided that $m_{\rm r}$ is by orders of magnitude less than the mass of the central body}, all the Keplerian orbital elements are left unaffected, apart from the mean anomaly $\mathcal{M}$. \textcolor{black}{The Lagrange equation for its variation  yields}
\eqi \dert{\mathcal{M}}t-n_{\rm b} = \rp{2Gm_{\rm r}}{n_{\rm b} a^3}.\eqf In it, $n_{\rm b} \doteq \sqrt{GM/a^3}$ is the Keplerian mean motion connected to the orbital period by $n_{\rm b}=2\pi/P_{\rm b}$.

The next non-vanishing averaged potential is $\left\langle U_2\right\rangle$,
which yields the following non-zero long-term variations
{\tiny{
\begin{align}\lb{microI}
\ang{\dert I t} & = \rp{3G\mg }{4n_{\rm b}\ton{1-\ee}^2   a^3}\ton{\rp{\rg}{a}}^2{\rm j_0}, \\ \nonumber \\
{\rm j0} \nonumber & = \sin \Ir \ton{\cos I \cos \Ir + \cos \Delta\Om \sin I \sin \Ir} \sin \Delta\Om, \\ \nonumber \\
\ang{\dert \Om t} & = -\rp{3G\mg }{16n_{\rm b}\ton{1-\ee}^2   a^3}\ton{\rp{\rg}{a}}^2{\rm h_0},\\ \nonumber \\
{\rm h0} \nonumber & = -2 \cos 2 I \cos \Delta\Om \csc I \sin 2 \Ir + \cos I \qua{\cos 2 \Ir \ton{3 + \cos 2 \Delta\Om} + 2 \sin^2 \Delta\Om}, \\ \nonumber \\
\ang{\dert \varpi t} & = \rp{3G\mg }{64n_{\rm b}\ton{1-\ee}^2   a^3}\ton{\rp{\rg}{a}}^2{\rm p_0},\\ \nonumber \\
{\rm p0} \nonumber & = \ton{3 - 4 \cos I + 5 \cos 2 I} \ton{1 + 3 \cos 2 \Ir} + 8 \ton{3 + 5 \cos I} \cos 2 \Delta\Om \sin^2 \ton{\rp{I}{2}} \sin^2 \Ir + \\ \nonumber \\
\nonumber & + 2 \cos \Delta\Om \sec\ton{\rp{I}{2}} \qua{\sin \ton{\rp{3}{2}I} + 5 \sin \ton{\rp{5}{2}I}} \sin 2 \Ir, \\ \nonumber \\
\ang{\dert{\mathcal{M}} t} - n_{\rm b} & = \rp{3G\mg }{64n_{\rm b}\ton{1-\ee}^{3/2}   a^3}\ton{\rp{\rg}{a}}^2{\rm g_0}, \\ \nonumber \\
{\rm g0} \nonumber & = \ton{1 + 3 \cos 2 I} \ton{1 + 3 \cos 2 \Ir} + 12 \cos 2 \Delta\Om \sin^2 I \sin^2 \Ir + 12 \cos \Delta\Om \sin 2 I \sin 2 \Ir;
\end{align}
}}
$\varpi\doteq\Omega +\omega$ is the longitude of pericenter.
As expected, all the perturbations vanish in the limit $R_{\rm r}\rightarrow 0$. If the ring and the planetary orbit share the same inclination to the reference $\{x,y\}$ plane, but not the same orientation, i.e. for $I=I_{\rm r},\ \Om \neq \Om_{\rm r}$, in general, non-vanishing perturbations depending on $I$ and $\Om - \Om_{\rm r}$ occur. In the case $\Om = \Om_{\rm r},\ I\neq I_{\rm r}$, the secular precession of the inclination $I$ vanishes, while
the rates of $\Om,\varpi,\mathcal{M}$ do not vanish, being functions of $I,I_{\rm r}$. Finally, if both the ring and the orbit lie in the same plane, i.e. for $I=I_{\rm r},\ \Om =\Om_{\rm r}$ (cfr. Table \ref{tavola}), then the inclination $I$ and the node $\Om$ remain unaffected, while the other non-zero precessions are
\begin{align}\lb{allora}
\ang{\dert\varpi t} & =  \rp{3Gm_{\rm r}}{4 a^3 \left(1 - e^2\right)^2 n_{\rm b}}\left(\rp{R_{\rm r}}{a}\right)^2, \\ \nonumber\\
\lb{allora2}\ang{\dert{\mathcal{M}} t} - n_{\rm b} & =  \rp{3Gm_{\rm r}}{4 a^3 \left(1 - e^2\right)^{3/2} n_{\rm b}}\left(\rp{R_{\rm r}}{a}\right)^2.
\end{align}
It can be noticed that \rfr{allora} and \rfr{allora2} are independent of the specific values of $I$ and $\Om$.
 %; the rate of $\varpi$
 %coincides with \rfr{rate} up to terms of order $\mathcal{O}(R_{\rm r}^4/r^4)$.
%
%

In the case of $\left\langle U_4 \right\rangle$, the resulting long-term rates of variation of the osculating Keplerian orbital elements for a generic ring-orbit configuration cannot be displayed in full since they are quite cumbersome; they are exact in the sense that no approximations in $e, I, \Ir,\Om,\Or$ were adopted. In general, they do not vanish, and explicitly depend on $I$ and $\DO$.
For different inclinations but equal displacements with respect to the reference $x$ direction $(I \neq \Ir,\Om = \Or)$, it turns out that all the long-term rates of change are non-zero, explicitly depending on $I$ and $\Ir$. In particular, the shifts in $e$ and $I$ are of order $\mathcal{O}(e)$ and $\mathcal{O}(e^2)$, respectively. The case in which the inclinations are the same and the planes are shifted apart $(I =\Ir,\Om\neq\Or)$ is more complicated.

Much simpler expressions are obtained for $I = I_{\rm r},\ \Om = \Om_{\rm r}$.
\begin{align}\lb{superd}
\left\langle\dert a t\right\rangle & =  0, \\ \nonumber \\
\left\langle\dert e t\right\rangle & =  0, \\ \nonumber \\
\left\langle\dert I t\right\rangle & =  0, \\ \nonumber \\
\left\langle\dert \Om t\right\rangle & =  0, \\ \nonumber \\
\lb{piccoloPERI}\left\langle\dert\varpi t\right\rangle & =  \rp{45 Gm_{\rm r}\left(1 + \rp{3}{4}e^2\right)}{32a^3\left(1 - e^2\right)^4 n_{\rm b}}\left(\rp{R_{\rm r}}{a}\right)^4, \\ \nonumber \\
\left\langle\dert{\mathcal{M}} t\right\rangle - n_{\rm b} & =  \rp{135 Gm_{\rm r}e^2}{128 a^3\left(1 - e^2\right)^{7/2}n_{\rm b}}\left(\rp{R_{\rm r}}{a}\right)^4.
\end{align}
Also in this case, the non-vanishing rates neither depend on $I$ nor on $\Om$.
%The pericenter precession of \rfr{superd} agrees with the term of order $\mathcal{O}\left(R^4_{\rm r}/a^4\right)$  in \rfr{rate}.
%\eqi
%\begin{array}{lll}
%\rp{32768 a^3\left(1 - e^2\right)^3 n_{\rm b}}{45 e Gm_{\rm r}}\left(\rp{a}{R_{\rm r}}\right)^4\left\langle\dert e t\right\rangle & = & \left[13 + %7\left( \cos 2I + 3 \cos I\cos I_{\rm r} +
%\cos 2I_{\rm r}\right) + \right.\\ \\
%
%& + & \left. 28 \cos 2\Delta\Om \sin^2 I \sin^2 I_{\rm r} + \right.\\ \\
%
%& + & \left. 28 \cos\left(\Om - \Om_{\rm r} \right) \sin 2I \sin 2I_{\rm r}\right]\times \\ \\
%
%& \times & \left\{\left(1 + 3 \cos 2I_{\rm r}\right) \sin^2 I \sin 2\omega + \right.\\ \\
%
%& + & \left.\left[\left(3 + \cos 2I\right) \cos 2\Delta\Om \sin^2 I_{\rm r} - \right.\right.\\ \\
%
%& - & \left.\left. 2 \cos\Om \cos\Om_{\rm r} \sin 2I \sin 2I_{\rm r}\right] \sin 2\omega + \right.\\ \\
%
%& + & \left. 4 \cos I \cos 2\omega \sin^2 I_{\rm r} \sin 2\Delta\Om -\right. \\ \\
%
%& - & \left. 4 \sin I \sin 2I_{\rm r} \left[\cos 2\omega \sin\left(\Om - \Om_{\rm r} \right) + \right.\right.\\ \\
%
%& + & \left.\left.\cos I \sin 2\omega \sin\Om\sin\Om_{\rm r}\right]\right\}.
%\end{array}
%\eqf
%
\subsection{The circumsolar ring with $R_{\rm r}=0.13\ {\rm au}$}
For the sake of simplicity, we will assume coplanarity between the ring and the perturbed planetary orbit;  \rfr{allora} and \rfr{piccoloPERI}
yield
\eqi
\ang{\dert\varpi t} = \rp{3Gm_{\rm r} }{4 a^3\left(1-e^2\right)^2 n_{\rm b}}\left(\rp{R_{\rm r}}{a}\right)^2\left[1 + \rp{15}{8}\rp{\left(1 + \rp{3}{4}e^2\right)}{\left(1-e^2\right)^2}\left(\rp{R_{\rm r}}{a}\right)^2\right]\lb{rate}.
\eqf

In the case of Mercury ($a_{\mercury}=0.387\ {\rm au}$), and for $R_{\rm r}=0.13\ {\rm au}$ \cite{rawal011}, \rfr{rate} yields
\eqi \left\langle \dert\varpi t \right\rangle = m_{\rm r}\ 3.08\times 10^{-20}\ {\rm mas\ cty^{-1}\ kg^{-1}},\eqf
where mas cty$^{-1}$ is a shorthand for milliarcseconds per century.
An upper bound on $m_{\rm r}$ can be obtained from the latest determinations $\Delta\dot\varpi_{\mercury}^{\rm (meas)}$ of the maximum allowed value for any unmodeled effect impacting the Hermean secular perihelion precession recently inferred by \cite{fienga011} from observations.
\cite{fienga011}, who did not model the Solar ring,  obtained
\eqi \Delta\dot\varpi_{\mercury}^{\rm (meas)} = 0.4\pm 0.6\ {\rm mas\ cty^{-1}}.\lb{fienga}\eqf
Actually, there is also a general relativistic effect, caused by the angular momentum of the Sun, which was not modeled by \cite{fienga011}: it is the gravitomagnetic precession of the perihelion \cite{lense18}, which is predicted to be as large as \eqi \dot\varpi_{\mercury}^{\rm (LT)} = -2\ {\rm mas\ cty^{-1}}\lb{ltrate}\eqf  by using the value of the Solar angular momentum by helioseismology \cite{pijpers98}.
In the case of Mercury, also the impact of the mismodeling in the Sun's even zonal harmonic $J_2$, modeled by \cite{fienga011}, must be taken into account. By assuming a $10\%$ uncertainty in it \cite{Roz011}, it is
\eqi \dot\varpi_{\mercury}^{(\delta J_2)} = 2.5\ {\rm mas\ cty^{-1}}\lb{j2rate}.\eqf
Thus, both  \rfr{ltrate} and \rfr{j2rate} must be subtracted from \rfr{fienga} in inferring an upper bound on $m_{\rm r}$. As a result, from
\eqi 0.5\ {\rm mas\ cty^{-1}} + m_{\rm r}\ 3.08\times 10^{-20}\ {\rm mas\ cty^{-1}\ kg^{-1}}\leq 1\ {\rm mas\ cty^{-1}}\eqf
we have
\eqi m_{\rm r} \leq 1.6\times 10^{19}\ {\rm kg}= 4.3\times 10^{-6}\ m_{\oplus}.\lb{boundmer}\eqf
It is 3708 times smaller than the upper bound by \cite{rawal011}.

One may legitimately wonder if the use of \rfr{rate} is accurate enough, given the relatively small distance of the Hermean orbit from the putative perturbing ring considered.
Actually, the answer is positive since \rfr{boundmer}, inserted in \rfr{piccoloPERI}, yields just $0.57$ mas cty$^{-1}$. Thus, we conclude that the truncation level chosen in the expansion of the annulus' potential yielding \rfr{rate} is fully adequate to deal with Mercury and a hypothetical perturbing massive ring with $R_{\rm r} = 0.13\ {\rm au}$, in view of the present-day level of accuracy in constraining the supplementary Hermean perihelion precession.
\subsection{The circumsolar ring with $R_{\rm r} = 0.02\ {\rm au}$}
The same approach for Mercury and $R_{\rm r}=0.02\ {\rm au}$ yields
\eqi m_{\rm r} \leq 8.3\times 10^{20}\ {\rm kg}= 1.4\times 10^{-4}\ m_{\oplus};\lb{boundino}\eqf
it is tighter by about one order of magnitude than the upper limit proposed  by \cite{rawal011}.
Note that \rfr{boundino} and \rfr{piccoloPERI} yield just $0.01$ mas cty$^{-1}$.
\subsection{The ring with $R_{\rm r}=1\ {\rm au}$}
The perihelion precession of Mars ($a_{\mars} = 1.524\ {\rm au}$) can  be used to effectively constrain the mass of a ring with $R_{\rm r}=1\ {\rm au}$. \cite{fienga011} obtained
\eqi\Delta\dot\varpi_{\mars}^{\rm (meas)}=-0.04\pm 0.15\ {\rm mas\ cty^{-1}}\lb{marte}\eqf
for the Arean supplementary perihelion precession. In the case of Mars, \rfr{rate} and \rfr{marte} yield
\eqi  m_{\rm r}\ 2.1\times 10^{-20}\ {\rm mas\ cty^{-1}\ kg^{-1}}\leq 0.11\ {\rm mas\ cty^{-1}},\lb{ujio1} \eqf
which implies
\eqi m_{\rm r} \leq 5.3\times 10^{18}\ {\rm kg}= 8.8\times 10^{-7}\ m_{\oplus}. \lb{ring1au}\eqf
Since the Arean Lense-Thirring precession  amounts to just $\dot\varpi_{\mars}^{\rm (LT)} = -0.03\ {\rm mas\ cty^{-1}}$, i.e. 5 times smaller than the present-day uncertainty \cite{fienga011} in determining the Arean supplementary perihelion, we did not include it in \rfr{ujio1}. The same holds for the mismodeling in the Sun's quadrupole moment, since $\dot\varpi_{\mars}^{(\delta J_2)} = 0.01\ {\rm mas\ cty^{-1}}$.

Also in this case, it turns out that the level of truncation chosen is adequate, given the current level of accuracy in constraining the supplementary Arean perihelion precession. Indeed, \rfr{ring1au}, inserted in \rfr{piccoloPERI}, yields $0.05$ mas cty$^{-1}$, which is 3 times smaller than 0.15 mas cty$^{-1}$.
\section{The outer ring: $r< R_{\rm r}$}\lb{esterno}
In the case $({\rm II})$ of \rfr{casi},  the calculations are, in general, more involved.

It turns out that, although $\left\langle U_0\right\rangle$ does not vanish, it does not contribute to the long-term variations of the osculating Keplerian orbital elements since it is just made of $m_{\rm r}$ and $R_{\rm r}$. \textcolor{black}{Indeed, it is $U_0 = -Gm_{\rm r}/R_{\rm r}$.}
Moreover, $\left\langle U_1\right\rangle=0$.
As far as $\left\langle U_2 \right\rangle$ is concerned, it contributes to the orbital changes with
{\tiny{
\begin{align}
\ang{\dert e t} \lb{prima1} & = \rp{15eG\mg\sqrt{1-\ee}}{32 n_{\rm b}\rg^3}{\rm E1},\\ \nonumber \\
{\rm E1} \nonumber & = \cos 2 \omega \ton{-4 \sin I \sin 2 \Ir \sin \Delta\Om + 4 \cos I \sin^2\Ir \sin 2 \Delta\Om} + \\ \nonumber \\
\nonumber & + \qua{\ton{1 + 3 \cos 2 \Ir} \sin I^2 + \ton{3 + \cos 2 I} \cos 2 \Delta\Om \sin^2\Ir -
2 \cos \Delta\Om \sin 2 I \sin 2 \Ir} \sin 2 \omega, \\ \nonumber \\
\ang{\dert I t} & = \rp{3G\mg}{8 n_{\rm b}\sqrt{1-\ee}\rg^3}\ton{2 {\rm I0} + {\rm I2}\ee},\\ \nonumber \\
{\rm I0} \nonumber & = \sin \Ir \ton{\cos I \cos \Ir + \cos \Delta\Om \sin I \sin \Ir} \sin \Delta\Om, \\ \nonumber \\
{\rm I2} \nonumber & = \ton{\cos I \cos \Ir + \cos \Delta\Om \sin I \sin \Ir} \qua{\ton{3 + 5 \cos 2 \omega} \sin \Ir \sin \Delta\Om -
5 \ton{\cos \Ir \sin I - \cos I \cos \Delta\Om \sin \Ir} \sin 2 \omega}, \\ \nonumber \\
\ang{\dert\Om t} & = -\rp{3G\mg\csc I}{8 n_{\rm b}\sqrt{1-\ee}\rg^3}\ton{2 {\rm O0} + {\rm O2}\ee},\\ \nonumber \\
{\rm O0} \nonumber & = \ton{\cos \Ir \sin I - \cos I \cos \Delta\Om \sin \Ir} \ton{\cos I \cos \Ir + \cos \Delta\Om \sin I \sin \Ir}, \\ \nonumber \\
{\rm O2} \nonumber & = \ton{\cos I \cos \Ir + \cos \Delta\Om \sin I \sin \Ir} \qua{-\ton{-3 + 5 \cos 2 \omega} \ton{\cos \Ir \sin I -
\cos I \cos \Delta\Om \sin \Ir} - 5 \sin \Ir \sin \Delta\Om \sin 2 \omega}, \\ \nonumber \\
\ang{\dert\varpi t} \lb{ratecompleto2} & = \rp{3G\mg}{128\sqrt{1-\ee}n_{\rm b}\rg^3}\ton{{\rm P0} + {\rm P2}\ee}, \\ \nonumber \\
{\rm P0} \nonumber & = 2 \grf{\ton{1 + 3 \cos 2 \Ir} \ton{3 + 20 \cos 2 \omega \sin^2\ton{\rp{I}{2}}} +
6 \cos 2 \Delta\Om \ton{2 - 2 \cos I +5 \cos 2 \omega} \sin^2\Ir + 5 \cos 2 I \ton{1 + 3 \cos 2 \Ir +
2 \cos 2 \Delta\Om \cos2 \omega \sin^2\Ir} + \right.\\ \nonumber \\
\nonumber & + \left. 40 \sin I \sin 2 \Ir \sin \Delta\Om \sin2 \omega +
\cos I \qua{4 \ton{1 + 3 \cos 2 \Ir} \ton{-1 + 5 \cos 2 \omega \sin^2\ton{\rp{I}{2}}} +40 \cos 2 \Delta\Om \sin^2\ton{\rp{I}{2}}
\sin^2\Ir -40 \sin^2\Ir \sin 2 \Delta\Om \sin2 \omega} + \right.\\ \nonumber \\
\nonumber & + \left. 4 \cos \Delta\Om \qua{3 + 5 \cos 2 I +
\cos I \ton{6 - 20 \cos^2\ton{\rp{I}{2}} \cos 2 \omega}} \sin2 \Ir \tan \ton{\rp{I}{2}}}, \\ \nonumber \\
{\rm P2} \nonumber & = -4 \ton{1 + 3 \cos 2 \Ir} \ton{-1 + 3 \cos I + 10 \cos 2 \omega \sin^2\ton{\rp{I}{2}}} -
16 \cos 2 \Delta\Om \ton{5 \cos^2\ton{\rp{I}{2}} \cos 2 \omega + 3 \sin^2\ton{\rp{I}{2}}} \sin^2\Ir + \\ \nonumber \\
\nonumber & + 8 \cos \Delta\Om \ton{-3 + 5 \cos 2 \omega} \sec\ton{\rp{I}{2}} \sin 3 \ton{\rp{I}{2}} \sin 2 \Ir -
20 \sec \ton{\rp{I}{2}} \ton{3 \sin \ton{\rp{I}{2}} + \sin 3 \ton{\rp{I}{2}}} \sin 2 \Ir \sin \Delta\Om \sin 2 \omega +\\ \nonumber \\
\nonumber & + 80 \cos^2\ton{\rp{I}{2}} \sin^2\Ir \sin 2 \Delta\Om \sin 2 \omega, \\ \nonumber \\
\ang{\dert{\mathcal{M}}t} - n_{\rm b} \lb{prima2} & = -\rp{G\mg}{256 n_{\rm b}\rg^3}\ton{{\rm M0} + {\rm M2}\ee }, \\ \nonumber \\
{\rm M0} \nonumber & = 4 \ton{1 + 3 \cos 2 \Ir} \ton{7 + 21 \cos 2 I + 30 \cos 2 \omega \sin^2 I} +
12 \cos 2 \Delta\Om \qua{10 \ton{3 + \cos 2 I} \cos 2 \omega + 28 \sin^2 I} \sin^2\Ir - \\ \nonumber \\
\nonumber & - 48 \cos \Delta\Om \ton{-7 + 5 \cos 2 \omega} \sin 2 I \sin 2 \Ir + 480 \sin I \sin 2 \Ir \sin \Delta\Om \sin 2 \omega -
480 \cos I \sin^2\Ir \sin 2 \Delta\Om \sin 2 \omega, \\ \nonumber \\
{\rm M2} \nonumber & = 6 \grf{2 \ton{1 + 3 \cos 2 \Ir} \ton{1 + 3 \cos 2 I + 10 \cos 2 \omega \sin^2 I} +
2 \cos 2 \Delta\Om \qua{10 \ton{3 + \cos 2 I} \cos 2 \omega + 12 \sin^2 I} \sin^2\Ir -\right.\\ \nonumber \\
\nonumber & - \left. 8 \cos \Delta\Om \ton{-3 + 5 \cos 2 \omega} \sin 2 I \sin 2 \Ir + 80 \sin I \sin 2 \Ir \sin \Delta\Om \sin
2 \omega - 80 \cos I \sin^2\Ir \sin 2 \Delta\Om \sin 2 \omega};
\end{align}
}}
the semimajor axis is left unaffected. As expected, all the perturbations of \rfr{prima1}-\rfr{prima2} vanish in the limit $R_{\rm r}\rightarrow \infty$. If the ring and the planetary orbit share the same inclination to the reference $\{x,y\}$ plane, but not the same orientation, i.e. for $I=I_{\rm r},\ \Om \neq \Om_{\rm r}$, in general, all the long-term changes of \rfr{prima1}-\rfr{prima2} do not vanish, depending on $I,\Delta\Om,\omega$; it is the case of, e.g., the three spacecrafts of LISA \cite{li08}.
If $\Om =\Om_{\rm r},\ I\neq I_{\rm r}$, then \rfr{prima1}-\rfr{prima2} are not zero, being functions of $I,I_{\rm r},\omega$.
 Finally, if both the ring and the orbit lie in the same plane, i.e. for $I=I_{\rm r},\ \Om =\Om_{\rm r}$ (cfr. Table \ref{tavola}), then  $e,I,\Om$ remain unaffected, while the long-term variations of $\varpi$ and $\mathcal{M}$ do not vanish, being equal to
\begin{align}\lb{dunque}
\ang{\dert\varpi t} & =  \rp{3\sqrt{1-e^2}Gm_{\rm r}}{4 n_{\rm b} R^3_{\rm r}}, \\ \nonumber \\
\lb{dunque2}\ang{\dert{\mathcal{M}} t} - n_{\rm b} & =  -\rp{7\left(1 + \rp{3}{7}e^2\right)Gm_{\rm r}}{4 n_{\rm b} R^3_{\rm r}}
\end{align}
Note that \rfr{dunque} and \rfr{dunque2} do not depend on the specific values of $I$ and $\Om$.
%the rate of $\varpi$  coincides with \rfr{rate2} up to terms of order $\mathcal{O}(R_{\rm r}^k/r^k),\ k\geq 2$.

The long-term variations of the Keplerian orbital elements due to $\ang{U_4}$ are quite involved for the general case $I\neq I_{\rm r},\ \Om \neq \Om_{\rm r}$: they cannot be explicitly displayed. Also in this case, no approximations in $e, I,\Ir$ were adopted. In the simpler coplanar case, i.e. for $I = I_{\rm r},\ \Om = \Om_{\rm r}$, they are
\begin{align}
\ang{\dert a t} & =  0, \\ \nonumber \\
\ang{\dert e t} & =  0, \\ \nonumber \\
\ang{\dert I t} & =  0, \\ \nonumber \\
\ang{\dert \Om t} & =  0, \\ \nonumber \\
\lb{superd2}\ang{\dert\varpi t} & =  \rp{45 Gm_{\rm r}\sqrt{1 - e^2}\left(1 + \rp{3}{4}e^2\right)}{32 n_{\rm b}R_{\rm r}^3}\left(\rp{a}{R_{\rm r}}\right)^2, \\ \nonumber \\
\lb{superd22}\ang{\dert{\mathcal{M}} t} - n_{\rm b} & =  -\rp{81 Gm_{\rm r}\left[1 + \rp{25}{12}e^2\left(1 + \rp{e^2}{5}\right)\right]}{32n_{\rm b} R_{\rm r}^3}\left(\rp{a}{R_{\rm r}}\right)^2.
\end{align}
Also in this case, the precessions of \rfr{superd2} and \rfr{superd22} are independent of $I$ and $\Om$.
%the perihelion rate agrees with \rfr{rate2} up to terms of order
%$\mathcal{O}\left(a^4/R_{\rm r}^4\right)$.
%

For other analytical calculation, see, e.g., \cite{kuchynka010}.
%analytically worked out approximate expressions for the long-term perturbations of the orbital elements of a planet due to an external ring by taking into %account its departures from circularity and coplanarity with respect to the reference $\{x,y\}$ plane by introducing annulus' eccentricity $e^{'}$ and %inclination $I^{'}$. They found non-vanishing effects for all the orbital elements of the perturbed planet, apart from its semimajor axis.   To lowest order in %the planet's eccentricity $e$ and inclination $I$,
%\cite{kuchynka010} found that a circular ($e^{'}=0$) and coplanar ($I^{'}=0$) matter ring induces non-vanishing secular perturbations on the longitude of %pericenter $\varpi$, the longitude of the ascending node $\Om$ and the mean longitude $\lambda\doteq \varpi + \mathcal{M}$ \cite[p. 34]{murray99}..
%
\subsection{The ring with $R_{\rm r}=1\ {\rm au}$}
As far as the perihelion precession in the coplanar case is concerned, \rfr{dunque} and \rfr{superd2}
yield
\eqi\ang{\dert\varpi t} =\rp{3G m_{\rm r}\sqrt{1-e^2}}{4 n_{\rm b} R^3_{\rm r}}\left[1 + \rp{15}{8}\left(1 + \rp{3}{4} e^2\right)\left(\rp{a}{R_{\rm r}}\right)^2\right].\lb{rate2}\eqf
 By proceeding with \rfr{fienga} as in the case of the inner ring,  a tight constrain can be obtained by using the perihelion of Venus ($a_{\venus}=0.723\ {\rm au}$), for which \cite{fienga011}
 obtained
 \eqi\Delta\dot\varpi_{\venus}^{\rm (meas)}= 0.2\pm 1.5\ {\rm mas\ cty^{-1}}.\lb{venere}\eqf
By neglecting the Cytherean unmodeled Lense-Thirring and mismodeled $J_2$ precessions $\dot\varpi_{\venus}^{\rm (LT)}=-0.3\ {\rm mas\ cty^{-1}},
\dot\varpi_{\venus}^{(\delta J_2)} = 0.2\ {\rm mas\ cty^{-1}}$,
\rfr{rate2} and \rfr{venere} yield
\eqi m_{\rm r}\leq 2.1\times 10^{19}\ {\rm kg}= 3.4\times 10^{-6}\ m_{\oplus}\lb{limite3}.\eqf
It must be noticed that, for $R_{\rm r}=1\ {\rm au}$, the Cytherean bound of  \rfr{limite3} is weaker than the Arean constraint of \rfr{ring1au}. Anyway, the use of \rfr{rate2} to deal with Venus and the hypothetical ring at $R_{\rm r} = 1\ {\rm au}$ is adequate, given the current level of accuracy in constraining the supplementary Cytherean perihelion precession. Indeed, \rfr{superd2}, computed with \rfr{limite3}, yields just $0.6$ mas cty$^{-1}$.
\subsection{The ring of minor asteroids between Mars and Jupiter}
The overall mass of the ring with which the dynamical action of the minor asteroids orbiting between Mars and Jupiter is usually modeled can effectively be constrained by using the perihelion of Mars and the result of \rfr{marte} for its supplementary precession \cite{fienga011}.
The impact of the individual largest perturbers of the asteroid belt on the motion of the inner planets, especially Mars and the Earth, was studied by, e.g., \cite{williams84,standish02,fienga05,mouret09,souchay09,somenzi010}. As far as their impact on some proposed spacecraft-based missions for fundamental physics are concerned, see, e.g., \cite{su99,vinet06,mouret011}.

By adopting \cite{krasinsky02,standish03,konopliv06,fienga08,kuchynka010} $R_{\rm r}=2.80\ {\rm au}$,
\rfr{rate2}, computed for Mars, yields
\eqi  m_{\rm r}\ 7.5\times 10^{-21}\ {\rm mas\ cty^{-1}\ kg^{-1}}\leq 0.11\ {\rm mas\ cty^{-1}},\lb{ujio} \eqf
from which it can be inferred
\eqi m_{\rm r} \leq 1.46\times 10^{19}\ {\rm kg}= 7.3\times 10^{-12}\ M_{\odot}. \lb{astmass}\eqf
Also in this case,  we did not consider the Arean Lense-Thirring effect in \rfr{ujio} because of its smallness.

Our result of \rfr{astmass} is smaller than the values obtained by \cite{krasinsky02} ($m_{\rm r}=(5\pm 1)\times 10^{-10}\ M_{\odot}$), \cite{standish03} ($m_{\rm r}=(6\pm 1)\times 10^{-11}\ M_{\odot}$), \cite{konopliv06} ($m_{\rm r}\lesssim 2.4\times 10^{-10}\ M_{\odot}$), \cite{fienga08} ($m_{\rm r}=(3.4\pm 1.5)\times 10^{-11}\ M_{\odot}$), and \cite{kuchynka09} ($m_{\rm r}=(6\pm 2)\times 10^{-11}\ M_{\odot}$).
In the case of \cite{pitjeva05a,fienga09,kuchynka09,pitjeva010b} $R_{\rm r}=3.14\ {\rm au}$, we obtain from Mars
\eqi m_{\rm r} \leq 2.33\times 10^{19}\ {\rm kg}= 1.1\times 10^{-11}\ M_{\odot}. \lb{astmass2}\eqf
\cite{kuchynka09} and \cite{fienga09} obtained $m_{\rm r}=(1.0\pm 0.3)\times 10^{-10}\ M_{\odot}$, which is one order of magnitude larger than \rfr{astmass2}. \cite{pitjeva05a} and \cite{pitjeva010b} obtained $m_{\rm r} = (3.35\pm 0.35)\times 10^{-10}\ M_{\odot}$ and $m_{\rm r} = (8.7\pm 3.5)\times 10^{-11}\ M_{\odot}$, respectively.

The use of \rfr{kalimba} and \rfr{super} for $r< R_{\rm r}$, yielding \rfr{dunque} and \rfr{superd2} in the coplanar case, might be questioned because of the closeness of the Arean orbit to the asteroid ring.
Actually, it is not the case, as it can be a-posteriori inferred from \rfr{astmass} and \rfr{astmass2}. Indeed, it turns out that \rfr{superd2}, computed with both \rfr{astmass}  ($R_{\rm r}=2.80\ {\rm au}$) and \rfr{astmass2} ($R_{\rm r} = 3.14\ {\rm au}$), yields just $0.03$ mas cty$^{-1}$, which is 5 times smaller than $0.15$ mas cty$^{-1}$.
Note that the same conclusion holds also if we adopt the upper limits of the figures for $m_{\rm r}$ obtained by other authors. Indeed, for $m_{\rm r} = 8\times 10^{-11}\ M_{\odot}$ \cite{kuchynka09} and $R_{\rm r} = 2.80\ {\rm au}$, \rfr{superd2} yields $0.11$ mas cty$^{-1}$. Moreover, $m_{\rm r} = 1.22\times 10^{-10}\ M_{\odot}$ and $R_{\rm r} = 3.14\ {\rm au}$ \cite{pitjeva010b}, inserted in \rfr{superd2}, give $0.07$ mas cty$^{-1}$.
Thus, we conclude that our expansion to $\alpha^4$, giving \rfr{rate2}, is fully adequate to treat the case of Mars perturbed by the ring of  minor asteroids within the present-day level of accuracy in constraining the supplementary precession of the Arean perihelion.
\subsection{The belt of the Trans-Neptunian Objects}
 Concerning the mass of the TNO belt, modeled as a circular ring with \cite{pitjeva010a,pitjeva010b} $R_{\rm r}=43\ {\rm au}$, it can be constrained with Saturn ($a_{\saturn}=9.582\ {\rm au}$) by using \cite{fienga011}
 \eqi\Delta\dot\varpi_{\saturn}^{\rm (meas)}=0.15\pm 0.65\ {\rm mas\ cty^{-1}}\eqf
 and \rfr{rate2}.
 Also in this case, there is no need to take into account the Kronian Lense-Thirring and $\delta J_2$ precessions since they are far smaller than $0.65$ mas cty$^{-1}$ ($\dot\varpi_{\saturn}^{\rm (LT)} = -1\times 10^{-4}\ {\rm mas\ cty^{-1}}, \dot\varpi_{\saturn}^{(\delta J_2)} = 2\times 10^{-5}\ {\rm mas\ cty^{-1}}$).
 Thus, from
 \eqi  m_{\rm r}\ 1.98\times 10^{-23}\ {\rm mas\ cty^{-1}\ kg^{-1}}\leq 0.8\ {\rm mas\ cty^{-1}}, \eqf
 it turns out
 \eqi m_{\rm r} \leq 4.03\times 10^{22}\ {\rm kg}= 2.0\times 10^{-8}\ M_{\odot}= 6.6\times 10^{-3}\ m_{\oplus}= 42.7\ m_{\rm Ceres}, \lb{tnomass}\eqf
 where we adopted \cite{carry08}
 \eqi m_{\rm Ceres} = 9.43\times 10^{20}\ {\rm kg}\eqf
 for the dwarf planet Ceres.

 For a comparison, according to \cite{pitjeva010a}, the minimum mass of the TNO belt would be as large as 110 masses of Ceres, while the maximum mass of the ring is expected to be 100 times the minimum mass, i.e. $1.1\times 10^4\ m_{\rm Ceres}$.  In other tests by \cite{pitjeva010a}, the maximum mass surpasses the minimum mass by 25, 50, and 75 times. Thus, the Kronian bound of \rfr{tnomass} is much more stringent than that by  \cite{pitjeva010a}.
 \cite{pitjeva010b} inferred for the mass of the TNO ring the smaller value $m_{\rm r}=(4.98\pm 0.14)\times 10^{-8}\ M_{\odot}=(105\pm 3)\ m_{\rm Ceres}$.
 The bound of \rfr{tnomass} is also much tighter than those obtained some years ago by \cite{iorio07} from earlier corrections $\Delta\dot\varpi^{\rm (meas)}$ to the standard precessions of the perihelia of the inner planets \cite{pitjeva05b}.

 We remark that, also in this case, the level of truncation considered is fully adequate, given the current level of accuracy in constraining the Kronian perihelion precession. Indeed, even by using the largest admissible value $m_{\rm r}=5.12\times 10^{-8}\ M_{\odot}$ by  \cite{pitjeva010b}, it turns out that \rfr{superd2} yields just $0.008$ mas cty$^{-1}$.
\section{Summary and conclusions}\lb{conclusioni}
We considered a test particle orbiting a central body of mass $M$ lying at the center of  a circular massive ring of radius $\rg$. We analytically worked out the long-term variations of the osculating Keplerian orbital elements $a,e,I,\Om,\varpi,\mathcal{M}$ of the particle induced by the first terms in the expansion of the gravitational potential $U_{\rm r}$ of the annulus. We considered both the case in which the particle moves outside ($r > \rg$) and inside ($r < \rg$) it. We did not restrict to any specific mutual orientation of the ring and the particle's orbital plane which, in general, neither coincide $(I\neq \Ir,\Om\neq\Or)$ nor lie in the  reference $\{x,y\}$ plane of the coordinate system adopted $(I\neq 0,\Ir\neq 0)$. In general, the resulting perturbations affect all the particle's orbital elements, apart from the semimajor axis $a$. They depend in a quite intricate way on the inclinations $I,\Ir$ and the nodes $\Om,\Or$ of both the particle and the ring, apart from the case in which they are coplanar $(I=\Ir,\Om=\Or)$.

By assuming coplanarity, we applied our results  to some hypothesized and existing annular distributions like circumsolar matter rings (Vulcanoids), a ring at 1 au, and the rings with which the dynamical actions  of the minor asteroids in the asteroid belt and of the Trans-Neptunian Objects are usually modeled. By comparing our analytical results to the latest observational determinations of the supplementary precessions $\Delta\dot\varpi^{(\rm meas)}$ of the longitudes of the perihelia $\varpi$ of some planets of the Solar System, we tentatively inferred upper bounds on the masses $m_{\rm r}$ of such massive annular distributions. As a complementary approach which could certainly be followed, a standard covariance-based analysis would consist of processing the same observational data records by explicitly modeling the  annular mass distributions whose mass one wants to determine/constrain, and treat it as a solve-for parameter to be estimated in the fit of the models to the observations. In any case, it should not be considered as the only legitimate method which must mandatorily be followed to obtain meaningful results. In general, the issue of potentially occurring mutual cancelation of various competing effects can meaningfully be inspected analytically as well, provided that one is able to work out proper expressions adequate to the accuracy level of the data. In this respect, it is not true that the effects caused by the rings are the same as for, say, the Sun's oblateness: indeed, their analytical expressions are quite different. Moreover, in most cases, several competing effects were actually modeled, so that they can impact the supplementary perihelion precessions only with their mismodeling, which can well be taken into account in evaluating their \virg{residual} effect. We stress that the constraints obtained by us cannot be considered as unrealistically tight; on the contrary, they might be regarded as relatively \virg{generous} since, in inferring them, we assumed just that the entire range of variation of the supplementary perihelion precessions is due to the dynamical action of the rings. We also tested the validity of such an assumption by checking in each case if competing mismodeled/unmodeled known effects like the Sun's $J_2$ and the Lense-Thirring effect were actually below the uncertainties released for the supplementary perihelion precessions; whenever it was not the case, like for Mercury, we took them into account  as well. Claiming a-priori that there might  still be room, in principle, for the action of any other unknown force conspiring in partly canceling allegedly larger ring effects  makes little sense. After all, the effect of any sort of \virg{Russell teacups} may well creep into the  the solved-for rings' masses estimated in a covariance analysis, if they really existed in Nature and they were not modeled at all. Moreover, we successfully checked that the level of approximation used for our formulas is adequate, given the present-day level of accuracy in constraining the supplementary precessions of the planetary perihelia. On the other hand, apart from the fact that our constraints are compatible with other, previously obtained bounds, it is certainly not strange that they are smaller then them, given that we adopted the latest ephemerides.

Our results are summarized in Table \ref{finale}.
\begin{table*}[ht!]
\caption{Upper bounds on the mass $m_{\rm r}$ of different circular matter rings of natural origin, whose radii $R_{\rm r}$ are listed in au, inferred from the corrections $\Delta\dot\varpi^{(\rm meas)}$ to the standard perihelion precessions of some planets of the Solar System by \cite{fienga011}. We used \rfr{rate} and \rfr{rate2} by assuming coplanarity between the rings and the planetary orbits. The upper bound for the TNOs ($R_{\rm r} = 43$ au) is equivalent
to $6.6\times 10^{-3}\ m_{\oplus}=42.7\ m_{\rm Ceres}$.
}\label{finale}
\centering
\bigskip
\begin{tabular}{llll}
\hline\noalign{\smallskip}
$\rg$ (au) & $m^{(\rm max)}_{\rm r}$ (kg) & $m^{(\rm max)}_{\rm r}$ ($m_{\oplus},\ M_{\odot}$) & Planet adopted \\
\noalign{\smallskip}\hline\noalign{\smallskip}
$0.02$ & $8.3\times 10^{20}$ &  $1.4\times 10^{-4}\ m_{\oplus}$ & Mercury \\
$0.13$ & $1.6\times 10^{19}$ &  $2.6\times 10^{-6}\ m_{\oplus}$ & Mercury \\
1 & $5.3 \times 10^{18}$ &   $8.8\times 10^{-7}\ m_{\oplus}$ & Mars \\
$2.80$ (minor asteroids) & $1.5 \times 10^{19}$ &  $7.3\times 10^{-12}\ M_{\odot}$ & Mars \\
$3.14$ (minor asteroids) & $2.3\times 10^{19}$ &  $1.1\times 10^{-11}\ M_{\odot}$ & Mars\\
$43$ (TNOs) & $4.0\times 10^{22}$ &  $2.0\times 10^{-8}\ M_{\odot}$ & Saturn \\
\noalign{\smallskip}\hline\noalign{\smallskip}
\end{tabular}
\end{table*}

In principle, our results are not limited just to baryonic matter distributions, being valid also for non-baryonic rings made of Dark Matter. Concerning putative circumsolar massive rings, which may, in principle, be made of non-baryonic Dark Matter as well, their mass is no larger than about
$ 1\times 10^{-4}-3\times 10^{-6}\ m_{\oplus}$, while the mass of a ring at $1\ {\rm au}$ is $m_{\rm r}\lesssim 9\times 10^{-7}\ m_{\oplus}$. Moreover, due their generality, our findings can be applied to different astronomical and astrophysical scenarios like Earth satellites, exoplanetary systems, stellar orbits around supermassive black holes, etc.
\section*{Acknowledgements}
I thank M. Efroimsky, US Naval Observatory, for having reviewed this manuscript.
\bibliography{Anellobib}{}

\providecommand{\href}[2]{#2}\begingroup\raggedright\begin{thebibliography}{10%
0}

\bibitem{miner07}
E.~D. Miner, R.~R. Wessen, and J.~N. Cuzzi, {\em Planetary Ring Systems}.
\newblock Springer Praxis Books, Chichester, 2007.

\bibitem{elliot77}
J.~L. Elliot, E.~Dunham, and D.~Mink, ``The rings of uranus,'' {\em Nature}
  {\bfseries 267} (May, 1977) 328--330.

\bibitem{johnson79}
T.~Johnson, ``Iau circular no. 3338,'' 1979.

\bibitem{smith79}
B.~A. Smith, L.~A. Soderblom, T.~V. Johnson, A.~P. Ingersoll, S.~A. Collins,
  E.~M. Shoemaker, G.~E. Hunt, H.~Masursky, M.~H. Carr, M.~E. Davies, A.~F.
  Cook, J.~M. Boyce, T.~Owen, G.~E. Danielson, C.~Sagan, R.~F. Beebe,
  J.~Veverka, J.~F. McCauley, R.~G. Strom, D.~Morrison, G.~A. Briggs, and V.~E.
  Suomi, ``The jupiter system through the eyes of voyager 1,'' {\em Science}
  {\bfseries 204} no.~4396, (1979) 951--957+960--972.

\bibitem{horn90}
L.~J. Horn, J.~Hui, A.~L. Lane, and J.~E. Colwell, ``Observations of neptunian
  rings by voyager photopolarimeter experiment,'' {\em Geophysical Research
  Letters} {\bfseries 17} (September, 1990) 1745--1748.

\bibitem{daubar99}
M.~E. Ockert-Bell, J.~A. Burns, I.~J. Daubar, P.~C. Thomas, J.~Veverka,
  M.~J.~S. Belton, and K.~P. Klaasen, ``The structure of jupiter's ring system
  as revealed by the galileo imaging experiment,'' {\em Icarus} {\bfseries 138}
  no.~2, (1999) 188--213.

\bibitem{porco04}
C.~C. Porco, R.~A. West, S.~Squyres, A.~McEwen, P.~Thomas, C.~D. Murray,
  A.~Delgenio, A.~P. Ingersoll, T.~V. Johnson, G.~Neukum, J.~Veverka, L.~Dones,
  A.~Brahic, J.~A. Burns, V.~Haemmerle, B.~Knowles, D.~Dawson, T.~Roatsch,
  K.~Beurle, and W.~Owen, ``Cassini imaging science: Instrument characteristics
  and anticipated scientific investigations at saturn,'' {\em Space Science
  Reviews} {\bfseries 115} no.~1-4, (2004) 363--497.

\bibitem{brecher79}
K.~Brecher, A.~Brecher, P.~Morrison, and I.~Wasserman, ``Is there a ring around
  the sun?,'' {\em Nature} {\bfseries 282} (November, 1979) 50--52.

\bibitem{rawal011}
J.~J. Rawal and S.~Ramadurai, ``Are there rings around the sun?,''
  \href{http://dx.doi.org/10.1007/s11038-011-9383-6}{{\em Earth Moon Planets}
  (2011) }.

\bibitem{krasinsky02}
G.~A. Krasinsky, E.~V. Pitjeva, M.~V. Vasilyev, and E.~I. Yagudina, ``Hidden
  mass in the asteroid belt,'' {\em Icarus} {\bfseries 158} no.~1, (2002)
  98--105.

\bibitem{edgeworth43}
K.~E. Edgeworth, ``The evolution of our planetary system,'' {\em Journal of the
  British Astronomical Association} {\bfseries 53} (July, 1943) 181--188.

\bibitem{kuiper51}
G.~P. Kuiper, ``On the origin of the solar system,'' in {\em Proceedings of a
  topical symposium, commemorating the 50th anniversary of the Yerkes
  Observatory and half a century of progress in astrophysics}, J.~A. Hynek,
  ed., p.~357.
\newblock McGraw-Hill, New York, 1951.

\bibitem{fernandez80}
J.~Fernandez, ``On the existence of a comet belt beyond neptune,'' {\em Monthly
  Notices of the Royal Astronomical Society} {\bfseries 192} (August, 1980)
  481--491.

\bibitem{pitjeva010a}
E.~V. Pitjeva, ``Epm ephemerides and relativity,'' in {\em Relativity in
  Fundamental Astronomy}, S.~A. Klioner, P.~K. Seidelmann, and M.~H. Soffel,
  eds., vol.~261 of {\em Proceedings IAU Symposium}, pp.~170--178.
\newblock Cambridge University Press, Cambridge, 2010.

\bibitem{pitjeva010b}
E.~V. Pitjeva, ``Influence of trans-neptunian objects on motion of major
  planets and limitation on the total tno mass from planet and spacecraft
  ranging,'' in {\em Icy Bodies of the Solar System}, J.~A. Fern\'{a}ndez,
  D.~Lazzaro, D.~Prialnik, and R.~Schulz, eds., vol.~263 of {\em Proceedings
  IAU Symposium}, pp.~93--97.
\newblock Cambridge University Press, Cambridge, 2010.

\bibitem{marshall011}
J.~P. Marshall, T.~L\"{o}hne, B.~Montesinos, A.~V. Krivov, C.~Eiroa, O.~Absil,
  G.~Bryden, J.~Maldonado, A.~Mora, J.~Sanz-Forcada, D.~Ardila, J.-C. Augereau,
  A.~Bayo, C.~Del~Burgo, W.~Danchi, S.~Ertel, D.~Fedele, M.~Fridlund,
  J.~Lebreton, B.~M. Gonz\'{a}lez-Garc\'{\i}a, R.~Liseau, G.~Meeus,
  S.~M\"{u}ller, G.~L. Pilbratt, A.~Roberge, K.~Stapelfeldt, P.~Th\'{e}bault,
  G.~J. White, and S.~Wolf, ``A herschel resolved far-infrared dust ring around
  hd 207129,'' {\em Astronomy and Astrophysics} {\bfseries 529} (May, 2011)
  A117.

\bibitem{vanleeuwen07}
F.~van Leeuwen, ``Validation of the new hipparcos reduction,'' {\em Astronomy
  and Astrophysics} {\bfseries 474} no.~2, (2007) 653--664.

\bibitem{poglitsch010}
A.~Poglitsch, C.~Waelkens, N.~Geis, H.~Feuchtgruber, B.~Vandenbussche,
  L.~Rodriguez, O.~Krause, E.~Renotte, C.~van Hoof, P.~Saraceno, J.~Cepa,
  F.~Kerschbaum, P.~Agn\`{e}se, B.~Ali, B.~Altieri, P.~Andreani, J.-L.
  Augueres, Z.~Balog, L.~Barl, O.~H. Bauer, N.~Belbachir, M.~Benedettini,
  N.~Billot, O.~Boulade, H.~Bischof, J.~Blommaert, E.~Callut, C.~Cara,
  R.~Cerulli, D.~Cesarsky, A.~Contursi, Y.~Creten, W.~De~Meester, V.~Doublier,
  E.~Doumayrou, L.~Duband, K.~Exter, R.~Genzel, J.-M. Gillis, U.~Gr\"{o}zinger,
  T.~Henning, J.~Herreros, R.~Huygen, M.~Inguscio, G.~Jakob, C.~Jamar, C.~Jean,
  J.~de~Jong, R.~Katterloher, C.~Kiss, U.~Klaas, D.~Lemke, D.~Lutz, S.~Madden,
  B.~Marquet, J.~Martignac, A.~Mazy, P.~Merken, F.~Montfort, L.~Morbidelli,
  T.~M\"{u}ller, M.~Nielbock, K.~Okumura, R.~Orfei, R.~Ottensamer, S.~Pezzuto,
  P.~Popesso, J.~Putzeys, S.~Regibo, V.~Reveret, P.~Royer, M.~Sauvage,
  J.~Schreiber, J.~Stegmaier, D.~Schmitt, J.~Schubert, E.~Sturm, M.~Thiel,
  G.~Tofani, R.~Vavrek, M.~Wetzstein, E.~Wieprecht, and E.~Wiezorrek, ``The
  photodetector array camera and spectrometer (pacs) on the herschel space
  observatory,'' {\em Astronomy and Astrophysics} {\bfseries 518} (July-august,
  2010) L2.

\bibitem{griffin010}
M.~J. Griffin, A.~Abergel, A.~Abreu, P.~A.~R. Ade, P.~André, J.-L. Augueres,
  T.~Babbedge, Y.~Bae, T.~Baillie, J.-P. Baluteau, M.~J. Barlow, G.~Bendo,
  D.~Benielli, J.~J. Bock, P.~Bonhomme, D.~Brisbin, C.~Brockley-Blatt,
  M.~Caldwell, C.~Cara, N.~Castro-Rodriguez, R.~Cerulli, P.~Chanial, S.~Chen,
  E.~Clark, D.~L. Clements, L.~Clerc, J.~Coker, D.~Communal, L.~Conversi,
  P.~Cox, D.~Crumb, C.~Cunningham, F.~Daly, G.~R. Davis, P.~de~Antoni,
  J.~Delderfield, N.~Devin, A.~di~Giorgio, I.~Didschuns, K.~Dohlen, M.~Donati,
  A.~Dowell, C.~D. Dowell, L.~Duband, L.~Dumaye, R.~J. Emery, M.~Ferlet,
  D.~Ferrand, J.~Fontignie, M.~Fox, A.~Franceschini, M.~Frerking, T.~Fulton,
  J.~Garcia, R.~Gastaud, W.~K. Gear, J.~Glenn, A.~Goizel, D.~K. Griffin,
  T.~Grundy, S.~Guest, L.~Guillemet, P.~C. Hargrave, M.~Harwit, P.~Hastings,
  E.~Hatziminaoglou, M.~Herman, B.~Hinde, V.~Hristov, M.~Huang, P.~Imhof, K.~J.
  Isaak, U.~Israelsson, R.~J. Ivison, D.~Jennings, B.~Kiernan, K.~J. King,
  A.~E. Lange, W.~Latter, G.~Laurent, P.~Laurent, S.~J. Leeks, E.~Lellouch,
  L.~Levenson, B.~Li, J.~Li, J.~Lilienthal, T.~Lim, S.~J. Liu, N.~Lu,
  S.~Madden, G.~Mainetti, P.~Marliani, D.~McKay, K.~Mercier, S.~Molinari,
  H.~Morris, H.~Moseley, J.~Mulder, M.~Mur, D.~A. Naylor, H.~Nguyen,
  B.~O'Halloran, S.~Oliver, G.~Olofsson, H.-G. Olofsson, R.~Orfei, M.~J. Page,
  I.~Pain, P.~Panuzzo, A.~Papageorgiou, G.~Parks, P.~Parr-Burman, A.~Pearce,
  C.~Pearson, I.~Pérez-Fournon, F.~Pinsard, G.~Pisano, J.~Podosek, M.~Pohlen,
  E.~T. Polehampton, D.~Pouliquen, D.~Rigopoulou, D.~Rizzo, I.~G. Roseboom,
  H.~Roussel, M.~Rowan-Robinson, B.~Rownd, P.~Saraceno, M.~Sauvage, R.~Savage,
  G.~Savini, E.~Sawyer, C.~Scharmberg, D.~Schmitt, N.~Schneider, B.~Schulz,
  A.~Schwartz, R.~Shafer, D.~L. Shupe, B.~Sibthorpe, S.~Sidher, A.~Smith, A.~J.
  Smith, D.~Smith, L.~Spencer, B.~Stobie, R.~Sudiwala, K.~Sukhatme, C.~Surace,
  J.~A. Stevens, B.~M. Swinyard, M.~Trichas, T.~Tourette, H.~Triou, S.~Tseng,
  C.~Tucker, A.~Turner, M.~Vaccari, I.~Valtchanov, L.~Vigroux, E.~Virique,
  G.~Voellmer, H.~Walker, R.~Ward, T.~Waskett, M.~Weilert, R.~Wesson, G.~J.
  White, N.~Whitehouse, C.~D. Wilson, B.~Winter, A.~L. Woodcraft, G.~S. Wright,
  C.~K. Xu, A.~Zavagno, M.~Zemcov, L.~Zhang, and E.~Zonca, ``The herschel-spire
  instrument and its in-flight performance,'' {\em Astronomy and Astrophysics}
  {\bfseries 518} (July-august, 2010) L3.

\bibitem{pilbratt010}
G.~L. Pilbratt, J.~R. Riedinger, T.~Passvogel, G.~Crone, D.~Doyle, U.~Gageur,
  A.~M. Heras, C.~Jewell, L.~Metcalfe, S.~Ott, and M.~Schmidt, ``Herschel space
  observatory. an esa facility for far-infrared and submillimetre astronomy,''
  {\em Astronomy and Astrophysics} {\bfseries 518} (July-august, 2010) L1.

\bibitem{nilsson010}
R.~Nilsson, R.~Liseau, A.~Brandeker, G.~Olofsson, G.~L. Pilbratt, C.~Risacher,
  J.~Rodmann, J.-C. Augereau, P.~Bergman, C.~Eiroa, M.~Fridlund,
  P.~Th\'{e}bault, and G.~J. White, ``Kuiper belts around nearby stars,'' {\em
  Astronomy and Astrophysics} {\bfseries 518} (July, 2010) A40.

\bibitem{shannon011}
A.~Shannon and Y.~Wu, ``Planetesimals in debris disks of sun-like stars,'' {\em
  The Astrophysical Journal} {\bfseries 739} no.~1, (2011) 36.

\bibitem{dezeeuw99}
P.~T. de~Zeeuw, R.~Hoogerwerf, J.~H.~J. de~Bruijne, A.~G.~A. Brown, and
  A.~Blaauw, ``A hipparcos census of the nearby ob associations,'' {\em The
  Astronomy Journal} {\bfseries 117} no.~1, (1999) 354--399.

\bibitem{gusten06}
R.~G\"{u}sten, L.~{\AA}. Nyman, P.~Schilke, K.~Menten, C.~Cesarsky, and
  R.~Booth, ``The atacama pathfinder experiment (apex)-a new submillimeter
  facility for southern skies-,'' {\em Astronomy and Astrophysics} {\bfseries
  454} no.~2, (2006) L13--L16.

\bibitem{nayakshin012}
S.~Nayakshin, S.~Sazonov, and R.~Sunyaev, ``Are supermassive black holes
  shrouded by 'super-oort' clouds of comets and asteroids?,'' {\em Monthly
  Notices of the Royal Astronomical Society} {\bfseries 419} no.~2, (2012)
  1238--1247.

\bibitem{ansorg05}
M.~Ansorg and D.~Petroff, ``Black holes surrounded by uniformly rotating
  rings,'' {\em Physical Review D} {\bfseries 72} no.~2, (2005) 024019.

\bibitem{karas04}
V.~Karas, J.-M. Hur\'{e}, and O.~Semer\'{a}k, ``Topical review: Gravitating
  discs around black holes,'' {\em Classical and Quantum Gravity} {\bfseries
  21} no.~7, (2004) R1--R51.

\bibitem{theys76}
J.~C. Theys and E.~A. Spiegel, ``Ring galaxies. i,'' {\em The Astrophysical
  Journal} {\bfseries 208} (September, 1976) 650--661.

\bibitem{theys77}
J.~C. Theys and E.~A. Spiegel, ``Ring galaxies. ii,'' {\em The Astrophysical
  Journal} {\bfseries 212} (March, 1977) 616--633.

\bibitem{whitmore90}
B.~Whitmore, R.~Lucas, D.~McElroy, T.~Steiman-Cameron, P.~Sackett, and
  R.~Olling, ``New observations and a photographic atlas of polar-ring
  galaxies,'' {\em The Astronomy Journal} {\bfseries 100} no.~5, (1990)
  1489--1522.

\bibitem{anz95}
P.~D. Anz-Meador, M.~J. Matney, and D.~J. Kessler, ``Physical properties of
  orbital debris and orbital debris clouds,'' {\em Advances in Space Research}
  {\bfseries 16} no.~11, (1995) 113--117.

\bibitem{flury95}
W.~Flury, ``The space debris environment of the earth,'' {\em Earth, Moon and
  Planets} {\bfseries 70} no.~1-3, (1995) 79--91.

\bibitem{kessler78}
D.~J. Kessler and B.~G. Cour-Palais, ``Collision frequency of artificial
  satellites: The creation of a debris belt,'' {\em Journal of Geophysical
  Research Space Physics} {\bfseries 83} no.~A6, (1978) 2637--2646.

\bibitem{montenbruck00}
O.~Montenbruck and E.~Gill, {\em Satellite Orbits. Models, Methods, and
  Applications}.
\newblock Springer-Verlag, Berlin, 2000.

\bibitem{liou08}
J.-C. Liou and N.~L. Johnson, ``Instability of the present leo satellite
  populations,'' {\em Advances in Space Research} {\bfseries 41} no.~7, (2008)
  1046--1053.

\bibitem{stansbery95}
E.~G. Stansbery, D.~J. Kessler, T.~E. Tracy, M.~J. Matney, and J.~F. Stanley,
  ``Characterization of the orbital debris environment from haystack radar
  measurements,'' {\em Advances in Space Research} {\bfseries 16} no.~11,
  (1995) 5--16.

\bibitem{friesen92}
L.~J. Friesen, A.~A. Jackson, H.~A. Zook, and D.~J. Kessler, ``Analysis of
  orbital perturbations acting on objects in orbits near geosynchronous earth
  orbit,'' {\em Journal of Geophysical Research Planets} {\bfseries 97} no.~E3,
  (1992) 3845--3863.

\bibitem{debris95}
N.~R.~C. Committee~on Space~Debris, {\em Orbital Debris: A Technical
  Assessment}.
\newblock The National Academies Press, 1995.
\newblock \url{http://www.nap.edu/openbook.php?record_id=4765}.

\bibitem{anselmo02}
L.~Anselmo and C.~Pardini, ``Collision risk mitigation in geostationary
  orbit,'' {\em Space Debris} {\bfseries 2} no.~2, (2002) 67--82.

\bibitem{klinkrad06}
H.~Klinkrad, {\em Space Debris. Models and Risk Analysis}.
\newblock Springer Praxis Books, Chichester, 2006.

\bibitem{poincare85a}
H.~Poincar\'{e}, ``Sur l'\'{e}quilibre d'une masse fluide anim\'{e}e d'un
  mouvement de rotation,'' {\em Comptes Rendus hebdomadaires des s\'{e}ances de
  l'Acad\'{e}mie des Sciences} {\bfseries 100} no.~January-June, (1885)
  346--348.

\bibitem{poincare85b}
H.~Poincar\'{e}, ``M\'{e}moires et observations. sur l'\'{e}quilibre d'une
  masse fluide anim\'{e}e d'un mouvement de rotation,'' {\em Bulletin
  Astronomique, Serie I} {\bfseries 2} (March, 1885) 109--118.

\bibitem{poincare85c}
H.~Poincar\'{e}, ``M\'{e}moires et observations. sur l'\'{e}quilibre d'une
  masse fluide anim\'{e}e d'un mouvement de rotation,'' {\em Bulletin
  Astronomique, Serie I} {\bfseries 2} (September, 1885) 405--413.

\bibitem{dyson92}
F.~W. Dyson, ``The potential of an anchor ring,'' {\em Philosophical
  Transactions of the Royal Society of London. Series A, Mathematical and
  Physical Sciences} {\bfseries 184} (January, 1893) 43--95.

\bibitem{dyson93}
F.~W. Dyson, ``The potential of an anchor ring. part ii,'' {\em Philosical
  Transactions of the Royal Society London A} {\bfseries 184} (January, 1893)
  1041--1106.

\bibitem{kowalewsky95}
S.~Kowalewsky, ``Zusatze und bermerkungen zu laplace's untersuchung uber die
  gestalt des saturnringes,'' {\em Astronomische Nachrichten} {\bfseries 111}
  no.~2643, (1885) 37--48.

\bibitem{lichtenstein33}
L.~Lichtenstein, {\em Gleichgewichtsfiguren rotierender Fl\"{u}ssigkeiten}.
\newblock Springer-Verlag, Berlin, 1933.

\bibitem{ansorg03a}
M.~Ansorg, A.~Kleinw\"{a}chter, and R.~Meinel, ``Uniformly rotating
  axisymmetric fluid configurations bifurcating from highly flattened maclaurin
  spheroids,'' {\em Monthly Notices of the Royal Astronomical Society}
  {\bfseries 339} no.~2, (2003) 515--523.

\bibitem{letelier07}
P.~S. Letelier, ``Simple potential-density pairs for flat rings,'' {\em Monthly
  Notices of the Royal Astronomical Society} {\bfseries 381} no.~3, (2007)
  1031--1034.

\bibitem{vogt09}
D.~Vogt and P.~S. Letelier, ``Analytical potential-density pairs for flat rings
  and toroidal structures,'' {\em Monthly Notices of the Royal Astronomical
  Society} {\bfseries 396} no.~3, (2009) 1487--1498.

\bibitem{morgan69}
T.~Morgan and L.~Morgan, ``The gravitational field of a disk,'' {\em Physical
  Review} {\bfseries 183} no.~5, (1969) 1097--1101.

\bibitem{kuzmin56}
G.~G. Kuzmin {\em Astron. Zh.} {\bfseries 33} (1956) 27.

\bibitem{toomre63}
A.~Toomre, ``On the distribution of matter within highly flattened galaxies,''
  {\em The Astrophysical Journal} {\bfseries 138} (1963) 385--392.

\bibitem{ciotti07}
L.~Ciotti and G.~Giampieri, ``Exact density-potential pairs from the
  holomorphic coulomb field,'' {\em Monthly Notices of the Royal Astronomical
  Society} {\bfseries 376} no.~3, (2007) 1162--1168.

\bibitem{ciotti08}
L.~Ciotti and F.~Marinacci, ``Exact density-potential pairs from
  complex-shifted axisymmetric systems,'' {\em Monthly Notices of the Royal
  Astronomical Society} {\bfseries 387} no.~3, (2008) 1117--1125.

\bibitem{petroff08}
D.~Petroff and S.~Horatschek, ``Uniformly rotating homogeneous and polytropic
  rings in newtonian gravity,'' {\em Monthly Notices of the Royal Astronomical
  Society} {\bfseries 389} no.~1, (2008) 156--172.

\bibitem{nishida94}
S.~Nishida and Y.~Eriguchi, ``A general relativistic toroid around a black
  hole,'' {\em The Astrophysical Journal} {\bfseries 427} no.~1, (1994)
  429--437.

\bibitem{vogt05}
D.~Vogt and P.~S. Letelier, ``Relativistic models of galaxies,'' {\em Monthly
  Notices of the Royal Astronomical Society} {\bfseries 363} no.~10, (2005)
  268--284.

\bibitem{ujevic011}
M.~Ujevic, P.~S. Letelier, and D.~Vogt, ``Relativistic ring models,'' {\em
  International Journal of Modern Physics D} {\bfseries 20} no.~12, (2011)
  2291--2304.

\bibitem{ansorg03b}
M.~Ansorg, A.~Kleinw\"{a}chter, and R.~Meinel, ``Relativistic dyson rings and
  their black hole limit,'' {\em The Astrophysical Journal Letters} {\bfseries
  582} no.~2, (2003) L87--L90.

\bibitem{horatschek010}
S.~Horatschek and D.~Petroff, ``Uniformly rotating homogeneous rings in
  post-newtonian gravity,'' {\em Monthly Notices of the Royal Astronomical
  Society} {\bfseries 408} no.~3, (2010) 1749--1757.

\bibitem{fischer05}
T.~Fischer, S.~Horatschek, and M.~Ansorg, ``Uniformly rotating rings in general
  relativity,'' {\em Monthly Notices of the Royal Astronomical Society}
  {\bfseries 364} no.~3, (2005) 943--947.

\bibitem{kleinwaechter011}
A.~Kleinw\"{a}chter, H.~Labranche, and R.~Meinel, ``On the black hole limit of
  rotating discs and rings,'' {\em General Relativity and Gravitation}
  {\bfseries 43} no.~5, (2011) 1469--1486.

\bibitem{kellog29}
O.~D. Kellog, {\em Foundations of Potential Theory}.
\newblock Dover, New York, 1929.

\bibitem{macmillan30}
W.~D. MacMillan, {\em The theory of the potential}.
\newblock McGraw Hill, New York, 1930.

\bibitem{abramowitz72}
M.~Abramowitz and I.~A. Stegun, eds., {\em Handbook of Mathematical Functions
  with Formulas, Graphs, and Mathematical Tables}.
\newblock Dover, New York, 9th printing~ed., 1972.

\bibitem{broucke05}
R.~A. Broucke and A.~Elipe, ``The dynamics of orbits in a potential field of a
  solid circular ring,'' {\em Regular and Chaotic Dynamics} {\bfseries 10}
  no.~2, (2005) 129--143.

\bibitem{Eck02}
D.~H. Eckhardt and J.~L.~G. Pesta\~{n}a, ``Technique for modeling the
  gravitational field of a galactic disk,'' {\em The Astrophysical Journal}
  {\bfseries 572} no.~2, (2002) L135--L137.

\bibitem{Zyp06}
F.~Zypman, ``Off-axis electric field of a ring of charge,'' {\em American
  Journal of Physics} {\bfseries 74} no.~4, (2006) 295--300.

\bibitem{Fuku010}
T.~Fukushima, ``Precise computation of acceleration due to uniform ring or
  disk,'' {\em Celestial Mechanics and Dynamical Astronomy} {\bfseries 108}
  no.~4, (2010) 339--356.

\bibitem{anderson02}
J.~D. Anderson, P.~A. Laing, E.~L. Lau, A.~S. Liu, M.~M. Nieto, and S.~G.
  Turyshev, ``Study of the anomalous acceleration of pioneer 10 and 11,'' {\em
  Physical Review D} {\bfseries 65} no.~8, (2002) 082004.

\bibitem{nieto05}
M.~M. Nieto, ``Analytic gravitational-force calculations for models of the
  kuiper belt, with application to the pioneer anomaly,'' {\em Physical Review
  D} {\bfseries 72} no.~8, (2005) 083004.

\bibitem{bertolami06}
O.~Bertolami and P.~Vieira, ``Pioneer anomaly and the kuiper belt mass
  distribution,'' {\em Classical and Quantum Gravity} {\bfseries 23} no.~14,
  (2006) 4625--4635.

\bibitem{dediego06}
J.~A. de~Diego, D.~N\'{u}\~{n}ez, and J.~Zavala, ``Pioneer anomaly?
  gravitational pull due to the kuiper belt,'' {\em International Journal of
  Modern Physics D} {\bfseries 15} no.~4, (2006) 533--544.

\bibitem{Vash012}
M.~A. Vashkovyak and S.~N. Vashkovyak, ``Force function of a slightly
  elliptical gaussian ring and its generalization to a nearly coplanar system
  of rings,'' {\em Solar System Reseach} {\bfseries 46} no.~1, (2012) 69--77.

\bibitem{iorio07}
L.~Iorio, ``Dynamical determination of the mass of the kuiper belt from motions
  of the inner planets of the solar system,'' {\em Monthly Notices of the Royal
  Astronomical Society} {\bfseries 375} no.~4, (2007) 1311--1314.

\bibitem{kuchynka010}
P.~Kuchynka, J.~Laskar, A.~Fienga, and H.~Manche, ``A ring as a model of the
  main belt in planetary ephemerides,'' {\em Astronomy and Astrophysics}
  {\bfseries 514} (May, 2010) A96.

\bibitem{khanna92}
R.~Khanna and S.~K. Chakrabarti, ``Effects of a self-gravitating disc on test
  particle motion around a kerr black hole,'' {\em Monthly Notices of the Royal
  Astronomical Society} {\bfseries 259} no.~1, (1992) 1--5.

\bibitem{sadeghian011}
L.~Sadeghian and C.~M. Will, ``Testing the black hole no-hair theorem at the
  galactic center: perturbing effects of stars in the surrounding cluster,''
  {\em Classical and Quantum Gravity} {\bfseries 28} no.~22, (2011) 225029.

\bibitem{fienga011}
A.~Fienga, J.~Laskar, P.~Kuchynka, H.~Manche, G.~Desvignes, M.~Gastineau,
  I.~Cognard, and G.~Theureau, ``The inpop10a planetary ephemeris and its
  applications in fundamental physics,'' {\em Celestial Mechanics and Dynamical
  Astronomy} {\bfseries 111} no.~3, (2011) 363--385.

\bibitem{murray99}
C.~D. Murray and S.~F. Dermott, {\em Solar System Dynamics}.
\newblock Cambridge University Press, Cambridge, 1999.

\bibitem{ni08}
W.-T. Ni, ``Astrod and astrod i — overview and progress,'' {\em International
  Journal of Modern Physics D} {\bfseries 17} no.~7, (2008) 921--940.

\bibitem{men010}
J.-R. Men, W.-T. Ni, and G.~Wang, ``Design of astrod-gw orbit,'' {\em Chinese
  Astronomy and Astrophysics} {\bfseries 34} no.~4, (2010) 434--446.

\bibitem{dong011}
P.~Dong and W.-T. Ni, ``Spacecraft gravitational wave detectors,'' {\em
  International Journal of Modern Physics D} {\bfseries 20} no.~10, (2011)
  2057--2062.

\bibitem{faller84}
J.~E. Faller and P.~L. Bender, ``A possible laser gravitational wave experiment
  in space,'' in {\em Precision measurement and fundamental constants II},
  B.~N. Taylor and W.~D. Phillips, eds., vol.~617 of {\em National Bureau of
  Standards Special Publication}, pp.~689--690.
\newblock United States. National Bureau of Standards, 1984.

\bibitem{faller85}
J.~E. Faller, P.~L. Bender, J.~L. Hall, D.~Hils, and M.~A. Vincent, ``Space
  antenna for gravitational wave astronomy,'' in {\em Proceedings of the
  Colloquium \virg{Kilometric Optical Arrays in Space}, Carg\`{e}se (Corsica)
  23-25 October 1984}, N.~Longdon and O.~Melita, eds., vol.~226 of {\em ESA
  Special Publication}, pp.~157--163.
\newblock 1985.

\bibitem{povoleri06}
A.~Povoleri and S.~Kemble, ``Lisa orbits,'' in {\em LASER INTERFEROMETER SPACE
  ANTENNA: Sixth International LISA Symposium}, S.~M. Merkowitz and J.~C.
  Livas, eds., vol.~873 of {\em AIP Conference Proceedings}, pp.~702--706.
\newblock American Institute of Physics, New York, 2006.

\bibitem{gath010}
P.~F. Gath, H.~R. Schulte, and D.~Weise, ``Challenges in the measurement and
  data-processing chain of the lisa mission,'' {\em Space Science Reviews}
  {\bfseries 151} no.~1-3, (2010) 61--73.

\bibitem{turon05}
C.~Turon, K.~S. O'Flaherty, and M.~A.~C. Perryman, eds., {\em The
  Three-Dimensional Universe with Gaia}.
\newblock ESA SP-576, 2005.

\bibitem{lindegren010}
L.~Lindegren, ``Gaia: Astrometric performance and current status of the
  project,'' in {\em Relativity in Fundamental Astronomy}, S.~A. Klioner, P.~K.
  Seidelmann, and M.~H. Soffel, eds., vol.~261 of {\em Proceedings IAU
  Symposium}, pp.~296--305.
\newblock Cambridge University Press, Cambridge, 2010.

\bibitem{Albe07}
A.~Alberti and C.~Vidal, ``Dynamics of a particle in a gravitational field of a
  homogeneous annulus disk celestial mechanics and dynamical astronomy,'' {\em
  Celestial Mechanics and Dynamical Astronomy} {\bfseries 98} no.~2, (2007)
  75--93.

\bibitem{Aze07a}
C.~Azev\^{e}do, H.~E. Cabral, and P.~Ontaneda, ``On the fixed homogeneous
  circle problem,'' {\em Advanced Nonlinear Studies} {\bfseries 7} no.~1,
  (2007) 47--75.

\bibitem{Aze07b}
C.~Azev\^{e}do and P.~Ontaneda, ``On the existence of periodic orbits for the
  fixed homogeneous circle problem,'' {\em Journal of Differential Equations}
  {\bfseries 235} no.~2, (2007) 341--365.

\bibitem{ramos011}
J.~Ramos-Caro, J.~F. Pedraza, and P.~S. Letelier, ``Motion around a monopole +
  ring system - i. stability of equatorial circular orbits versus regularity of
  three-dimensional motion,'' {\em Monthly Notices of the Royal Astronomical
  Society} {\bfseries 414} no.~4, (2011) 3105--3116.

\bibitem{tresaco011}
E.~Tresaco, A.~Elipe, and A.~Riaguas, ``Dynamics of a particle under the
  gravitational potential of a massive annulus: properties and equilibrium
  description,'' {\em Celestial Mechanics and Dynamical Astronomy} {\bfseries
  111} no.~4, (2011) 431--447.

\bibitem{tresaco012}
E.~Tresaco, A.~Elipe, and A.~Riaguas, ``Computation of families of periodic
  orbits and bifurcations around a massive annulus,'' {\em Astrophysics and
  Space Science} {\bfseries 338} no.~1, (2012) 23--33.

\bibitem{leverrier59}
U.-J. Le~Verrier, ``Lettre de m. le verrier \`{a} m. faye sur la th\'{e}orie de
  mercure et sur le mouvement du p\'{e}rih\'{e}lie de cette plan\`{e}te,'' {\em
  Comptes rendus hebdomadaires des s\'{e}ances de l'Acad\'{e}mie des sciences}
  {\bfseries 49} (July-december, 1859) 379--383.

\bibitem{sandage00}
T.~Sandage, {\em The Neptune File}.
\newblock The Penguin Press, New York, 2000.

\bibitem{newcomb82}
S.~Newcomb, ``Discussion and results of observations on transits of mercury,''
  in {\em Astronomical Papers of the American Ephemeris and Nautical Almanac},
  vol.~1, pp.~363--487.
\newblock U.S. Nautical Almanac Office, Washington, 1882.

\bibitem{nobili86}
A.~M. Nobili and C.~M. Will, ``The real value of mercury's perihelion
  advance,'' {\em Nature} {\bfseries 320} (March, 1986) 39--41.

\bibitem{seeliger06}
H.~von Seeliger, ``Das zodiakallicht und die empirischen glieder in der
  bewegung der innern planeten,'' {\em Sitzungsber. K\"{o}n. Bayer. Akad. Wiss.
  (M\"{u}nchen)} {\bfseries 36} (1906) 595--622.

\bibitem{poincare53}
H.~Poincar\'{e}, ``Les limites de la loi de newton,'' {\em Bulletin
  Astronomique} {\bfseries 17} (1953) chapter IV, p. 138; Conclusion, p. 265.

\bibitem{einstein15}
A.~Einstein, ``Erkl\"{a}rung der perihelbewegung des merkur aus der allgemeinen
  relativit\"{a}tstheorie,'' {\em Sitzungsberichte der Preu{\ss}ischen Akademie
  der Wissenschaften} {\bfseries 47} (November, 1915) 831--839.

\bibitem{courten72}
H.~C. Courten, ``Neue sonnennahe planetoiden,'' {\em Umschau Wiss. Tech.}
  {\bfseries 72} (1972) 562.

\bibitem{seargent011}
D.~A.~J. Seargent, {\em Weird Astronomy}.
\newblock Springer-Verlag, New York, 2011.
\newblock p. 68.

\bibitem{evans99}
N.~W. Evans and S.~Tabachnik, ``Possible long-lived asteroid belts in the inner
  solar system,'' {\em Nature} {\bfseries 399} no.~6731, (1999) 41--43.

\bibitem{lebofsky75}
L.~A. Lebofsky, ``Stability of frosts in the solar system,'' {\em Icarus}
  {\bfseries 25} no.~2, (1975) 205--217.

\bibitem{vokrouhlicky99}
D.~Vokrouhlick\'{y}, ``A complete linear model for the yarkovsky thermal force
  on spherical asteroid fragments,'' {\em Astronomy and Astrophysics}
  {\bfseries 344} (April, 1999) 362--366.

\bibitem{durda00}
D.~D. Durda, S.~A. Stern, W.~B. Colwell, J.~W. Parker, H.~F. Levison, and D.~M.
  Hassler, ``A new observational search for vulcanoids in soho/lasco
  coronagraph images,'' {\em Icarus} {\bfseries 148} no.~1, (2000) 312--315.

\bibitem{nottale97}
L.~Nottale, G.~Schumacher, and J.~Gay, ``Scale relativity and quantization of
  the solar system,'' {\em Astronomy and Astrophysics} {\bfseries 322} (June,
  1997) 1018--1025.

\bibitem{schumacher01}
G.~Schumacher and J.~Gay, ``An attempt to detect vulcanoids with soho/lasco
  images. i. scale relativity and quantization of the solar system,'' {\em
  Astronomy and Astrophysics} {\bfseries 368} no.~3, (2001) 1108--1114.

\bibitem{brueckner95}
G.~E. Brueckner, R.~A. Howard, M.~J. Koomen, C.~M. Korendyke, D.~J. Michels,
  J.~D. Moses, D.~G. Socker, K.~P. Dere, P.~L. Lamy, A.~Llebaria, M.~V. Bout,
  R.~Schwenn, G.~M. Simnett, D.~K. Bedford, and C.~J. Eyles, ``The large angle
  spectroscopic coronagraph (lasco),'' {\em Solar Physics} {\bfseries 162}
  no.~1-2, (1995) 357--402.

\bibitem{domingo95}
V.~Domingo, B.~Fleck, and A.~I. Poland, ``The soho mission: an overview,'' {\em
  Solar Physics} {\bfseries 162} no.~1-2, (1995) 1--37.

\bibitem{belton66}
M.~J.~S. Belton, ``Dynamics of interplanetary dust,'' {\em Science} {\bfseries
  151} no.~3706, (1966) 35--44.

\bibitem{belton67}
M.~J.~S. Belton, ``Dynamics of interplanetary dust particles near the sun,'' in
  {\em The Zodiacal Light and the Interplanetary Medium}, J.~L. Weinberg and
  H.~M. Mann, eds., no.~150 in NASA Special Publication.
\newblock National Technical Information Service, Springfield, Virginia, 1967.
\newblock p. 301.

\bibitem{kimura98}
H.~Kimura and I.~Mann, ``Brightness of the solar f-corona,'' {\em Earth,
  Planets and Space} {\bfseries 50} no.~6-7, (1998) 493--499.

\bibitem{singh04}
J.~Singh, T.~Sakurai, K.~Ichimoto, M.~Hagino, and T.~T. Yamamoto, ``Existence
  of nanoparticle dust grains in the inner solar corona?,'' {\em The
  Astrophysical Journal} {\bfseries 608} no.~1, (2004) L69--L72.

\bibitem{macqueen68}
R.~M. MacQueen, ``Infrared observations of the outer solar corona,'' {\em The
  Astrophysical Journal} {\bfseries 154} (December, 1968) 1059--1076.

\bibitem{peterson67}
A.~W. Peterson, ``Experimental detection of thermal radiation from
  interplanetary dust,'' {\em The Astrophysical Journal} {\bfseries 148}
  (April, 1967) L37--L39.

\bibitem{peterson69}
A.~W. Peterson, ``The coronal brightness at 2.23 microns,'' {\em The
  Astrophysical Journal} {\bfseries 155} (March, 1969) 1009--1016.

\bibitem{rao81}
U.~R. Rao, T.~K. Alex, V.~S. Iyengar, K.~Kasturirangan, T.~M.~K. Marar, R.~S.
  Mathur, and D.~P. Sharma, ``Ir observations of the solar corona - a ring
  around the sun,'' {\em Nature} {\bfseries 289} (February, 1981) 779--780.

\bibitem{mizutani84}
K.~Mizutani, T.~Maihara, N.~Hiromoto, and H.~Takami, ``Near-infrared
  observation of the circumsolar dust emission during the 1983 solar eclipse,''
  {\em Nature} {\bfseries 312} (November, 1984) 134--136.

\bibitem{hodapp92}
K.-W. Hodapp, R.~M. MacQueen, and D.~N.~B. Hall, ``A search during the 1991
  solar eclipse for the infrared signature of circumsolar dust,'' {\em Nature}
  {\bfseries 355} (Febraury, 1992) 707--710.

\bibitem{lamy92}
P.~Lamy, J.~R. Kuhn, H.~Lin, S.~Koutchmy, and R.~N. Smartt, ``No evidence of a
  circumsolar dust ring from infrared observations of the 1991 solar eclipse,''
  {\em Science} {\bfseries 257} no.~5075, (1992) 1377--1380.

\bibitem{debi95}
C.~Debi~Prasad, ``Variability of circumsolar dust ring,'' {\em Solar Physics}
  {\bfseries 159} no.~1, (1995) 181--190.

\bibitem{mann05}
I.~Mann and E.~Murad, ``On the existence of silicon nanodust near the sun,''
  {\em The Astrophysical Journal} {\bfseries 624} no.~2, (2005) L125--L128.

\bibitem{drell65}
S.~L. Drell, H.~M. Foley, and M.~A. Ruderman, ``Drag and propulsion of large
  satellites in the ionosphere: An alfv\'{e}n propulsion engine in space,''
  {\em Journal of Geophysical Research} {\bfseries 70} no.~13, (1965)
  3131--3145.

\bibitem{leinert07}
C.~Leinert and B.~Moster, ``Evidence for dust accumulation just outside the
  orbit of venus,'' {\em Astronomy and Astrophysics} {\bfseries 472} no.~1,
  (2007) 335--340.

\bibitem{desai81}
U.~D. Desai, ``Morphological study of the time histories of gamma-ray bursts,''
  {\em Astrophysics and Space Science} {\bfseries 75} no.~1, (1981) 15--20.

\bibitem{jackson89}
A.~A. Jackson and H.~A. Zook, ``A solar system dust ring with earth as its
  shepherd,'' {\em Nature} {\bfseries 337} (February, 1989) 629--631.

\bibitem{dermott94}
S.~F. Dermott, S.~Jayraraman, Y.~L. Xu, B.~{\AA}.~S. Gustafson, and J.~C. Liou,
  ``A circumsolar ring of asteroidal dust in resonant lock with the earth,''
  {\em Nature} {\bfseries 369} no.~6483, (1994) 719--723.

\bibitem{reach94}
W.~T. Reach, B.~A. Franz, J.~L. Weiland, M.~G. Hauser, T.~N. Kelsall, E.~L.
  Wright, G.~Rawleit, S.~W. Stemwedel, and W.~J. Spiesman, ``Observational
  confirmation of a circumsolar dust ring by the cobe satellite,'' {\em Nature}
  {\bfseries 374} (April, 1994) 521--523.

\bibitem{hauser93}
M.~G. Hauser, ``Cobe/dirbe observations of infrared emission from stars and
  dust,'' in {\em Back to the Galaxy}, S.~S. Holt and F.~Verter, eds., vol.~278
  of {\em AIP Conference Proceedings}, pp.~201--205.
\newblock American Institute of Physics, New York, 1993.

\bibitem{cobe}
N.~W. Boggess, J.~C. Mather, R.~Weiss, C.~L. Bennett, E.~S. Cheng, E.~Dwek,
  S.~Gulkis, M.~G. Hauser, M.~A. Janssen, T.~Kelsall, S.~S. Meyer, S.~H.
  Moseley, T.~L. Murdock, R.~A. Shafer, R.~F. Silverberg, G.~F. Smoot, D.~T.
  Wilkinson, and E.~L. Wright, ``The cobe mission-its design and performance
  two years after launch,'' {\em The Astrophysical Journal} {\bfseries 397}
  no.~2, (1992) 420--429.

\bibitem{reach010}
W.~T. Reach, ``Structure of the earth's circumsolar dust ring,'' {\em Icarus}
  {\bfseries 209} no.~2, (2010) 848--850.

\bibitem{mann06}
I.~Mann, M.~K\"{o}hler, H.~Kimura, A.~Cechowski, and T.~Minato, ``Dust in the
  solar system and in extra-solar planetary systems,'' {\em The Astronomy and
  Astrophysics Review} {\bfseries 13} no.~3, (2006) 159--228.

\bibitem{standish03}
E.~M. Standish, ``Jpl planetary ephemeris de409,'' Interoffice memorandum IOM
  312.N-03-007, 2003.

\bibitem{konopliv06}
A.~S. Konopliv, C.~F. Yoder, E.~M. Standish, D.-N. Yuan, and W.~L. Sjogren, ``A
  global solution for the mars static and seasonal gravity, mars orientation,
  phobos and deimos masses, and mars ephemeris,'' {\em Icarus} {\bfseries 182}
  no.~1, (2006) 23--50.

\bibitem{fienga08}
A.~Fienga, A.~Manche, J.~Laskar, and M.~Gastineau, ``Inpop06: a new numerical
  planetary ephemeris,'' {\em Astronomy and Astrophysics} {\bfseries 477}
  no.~1, (2008) 315--327.

\bibitem{pitjeva05a}
E.~V. Pitjeva, ``High-precision ephemerides of planets-epm and determination of
  some astronomical constants,'' {\em Solar System Research} {\bfseries 39}
  no.~3, (2005) 176--186.

\bibitem{fienga09}
A.~Fienga, J.~Laskar, T.~Morley, H.~Manche, P.~Kuchynka, C.~Le~Poncin-Lafitte,
  F.~Budnik, M.~Gastineau, and L.~Somenzi, ``Inpop08, a 4-d planetary
  ephemeris: from asteroid and time-scale computations to esa mars express and
  venus express contributions,'' {\em Astronomy and Astrophysics} {\bfseries
  507} no.~3, (2009) 1675--1686.

\bibitem{kuchynka09}
P.~Kuchynka, J.~Laskar, A.~Fienga, H.~Manche, and L.~Somenzi, ``Improving the
  asteroid perturbations modelling in planetary ephemerides,'' in {\em
  Proceedings of the \virg{Journ\'{e}es 2008 Syst\`{e}mes de r\'{e}f\'{e}rence
  spatio-temporels}}, M.~H. Soffel and N.~Capitaine, eds., pp.~84--85.
\newblock Lohrmann-Observatorium and Observatoire de Paris, Dresden, 2009.

\bibitem{zwicky33}
F.~Zwicky, ``Die rotverschiebung von extragalaktischen nebeln,'' {\em Helvetica
  Phisica Acta} {\bfseries 6} (1933) 110--127.

\bibitem{bosma81}
A.~Bosma, ``21-cm line studies of spiral galaxies. i-observations of the
  galaxies ngc 5033, 3198, 5055, 2841, and 7331. ii-the distribution and
  kinematics of neutral hydrogen in spiral galaxies of various morphological
  types,'' {\em The Astronomy Journal} {\bfseries 86} no.~12, (1981)
  1791--1846.

\bibitem{rubin82}
V.~C. Rubin, W.~K. Ford~Jr., N.~Thonnard, and D.~Burstein, ``Rotational
  properties of 23 sb galaxies,'' {\em The Astrophysical Journal} {\bfseries
  261} (October, 1982) 439--456.

\bibitem{zwicky37}
F.~Zwicky, ``On the masses of nebulae and of clusters of nebulae,'' {\em The
  Astrophysical Journal} {\bfseries 86} no.~3, (1937) 217--246.

\bibitem{rubin83}
V.~C. Rubin, ``The rotation of spiral galaxies,'' {\em Science} {\bfseries 220}
  no.~4604, (1983) 1339--1344.

\bibitem{chiu03}
W.~A. Chiu, X.~Fan, and J.~P. Ostriker, ``Combining wilkinson microwave
  anisotropy probe and sloan digital sky survey quasar data on reionization
  constrains cosmological parameters and star formation efficiency,'' {\em The
  Astrophysical Journal} {\bfseries 599} no.~2, (2003) 759--772.

\bibitem{komatsu09}
E.~Komatsu, J.~Dunkley, M.~R. Nolta, C.~L. Bennett, B.~Gold, G.~Hinshaw,
  N.~Jarosik, D.~Larson, M.~Limon, L.~Page, D.~N. Spergel, M.~Halpern, R.~S.
  Hill, A.~Kogut, S.~S. Meyer, G.~S. Tucker, J.~L. Weiland, E.~Wollack, and
  E.~L. Wright, ``Five-year wilkinson microwave anisotropy probe observations:
  Cosmological interpretation,'' {\em The Astrophysical Journal Supplement}
  {\bfseries 180} no.~2, (2009) 330--376.

\bibitem{tytler00}
D.~Tytler, J.~M. O'Meara, N.~Suzuki, and D.~Lubin, ``Review of big bang
  nucleosynthesis and primordial abundances,'' {\em Physica Scripta} {\bfseries
  2000} no.~T85, (2000) 12--31.

\bibitem{hinshaw09}
G.~Hinshaw, J.~L. Weiland, R.~S. Hill, N.~Odegard, D.~Larson, C.~L. Bennett,
  J.~Dunkley, B.~Gold, M.~R. Greason, N.~Jarosik, E.~Komatsu, M.~R. Nolta,
  L.~Page, D.~N. Spergel, E.~Wollack, M.~Halpern, A.~Kogut, M.~Limon, S.~S.
  Meyer, G.~S. Tucker, and E.~L. Wright, ``Five-year wilkinson microwave
  anisotropy probe observations: Data processing, sky maps, and basic
  results,'' {\em The Astrophysical Journal Supplement} {\bfseries 180} no.~2,
  (2009) 225--245.

\bibitem{steigman85}
G.~Steigman and M.~S. Turner, ``Cosmological constraints on the properties of
  weakly interacting massive particles,'' {\em Nuclear Physics B} {\bfseries
  253} (1985) 375--386.

\bibitem{jungman96}
G.~Jungman, M.~Kamionkowski, and K.~Griest, ``Supersymmetric dark matter,''
  {\em Physics Reports} {\bfseries 267} no.~5-6, (1996) 195--373.

\bibitem{haber85}
H.~E. Haber and G.~L. Kane, ``The search for supersymmetry: Probing physics
  beyond the standard model,'' {\em Physics Reports} {\bfseries 117} no.~2-4,
  (1985) 75--263.

\bibitem{gould91}
A.~Gould, ``Gravitational diffusion of solar system wimps,'' {\em The
  Astrophysical Journal} {\bfseries 368} (February, 1991) 610--615.

\bibitem{lundberg04}
J.~Lundberg and J.~Edsj\"{o}, ``Weakly interacting massive particle diffusion
  in the solar system including solar depletion and its effect on earth capture
  rates,'' {\em Physical Review D} {\bfseries 69} no.~12, (2004) 123505.

\bibitem{peter09a}
A.~H.~G. Peter, ``Dark matter in the solar system. iii. the distribution
  function of wimps at the earth from gravitational capture,'' {\em Physical
  Review D} {\bfseries 79} no.~10, (2009) 103533.

\bibitem{bertone05}
G.~Bertone, D.~Hooper, and J.~Silk, ``Particle dark matter: evidence,
  candidates and constraints,'' {\em Physics Report} {\bfseries 405} no.~5-6,
  (2005) 279--390.

\bibitem{khriplovich09}
I.~B. Khriplovich and D.~L. Shepelyansky, ``Capture of dark matter by the solar
  system,'' {\em International Journal of Modern Physics D} {\bfseries 18}
  no.~12, (2009) 1903--1912.

\bibitem{khriplovich010}
I.~B. Khriplovich, ``Comment on \virg{Comments on recent work on dark-matter
  capture in the Solar System},''
  \href{http://arxiv.org/abs/1005.1778}{{\ttfamily arXiv:1005.1778
  [astro-ph.EP]}}.

\bibitem{khriplovich011}
I.~B. Khriplovich, ``Capture of dark matter by the solar system:. simple
  estimates,'' {\em International Journal of Modern Physics D} {\bfseries 20}
  no.~1, (2011) 17--22.

\bibitem{edsjo010}
J.~Edsj\"{o} and A.~H.~G. Peter, ``Comments on recent work on dark-matter
  capture in the solar system,''
  \href{http://arxiv.org/abs/1004.5258}{{\ttfamily arXiv:1004.5258
  [astro-ph.EP]}}.

\bibitem{damour98}
T.~Damour and L.~M. Krauss, ``A new solar system dark matter population of
  weakly interacting massive particles,'' {\em Physical Review Letters}
  {\bfseries 81} no.~26, (1998) 5726--5729.

\bibitem{gould88}
A.~Gould, ``Direct and indirect capture of weakly interacting massive particles
  by the earth,'' {\em The Astrophysical Journal} {\bfseries 328} (May, 1988)
  919--939.

\bibitem{kozai62}
Y.~Kozai, ``Secular perturbations of asteroids with high inclination and
  eccentricity,'' {\em The Astronomy Journal} {\bfseries 67} no.~9, (1962)
  591--598.

\bibitem{damour99}
T.~Damour and L.~M. Krauss, ``New wimp population in the solar system and new
  signals for dark-matter detectors,'' {\em Physical Review D} {\bfseries 59}
  no.~6, (1999) 063509.

\bibitem{peter09b}
A.~H.~G. Peter, ``Dark matter in the solar system. i. the distribution function
  of wimps at the earth from solar capture,'' {\em Physical Review D}
  {\bfseries 79} no.~10, (2009) 103531.

\bibitem{akerib03}
D.~S. Akerib, J.~Alvaro-Dean, M.~S. Armel, M.~J. Attisha, L.~Baudis, D.~A.
  Bauer, A.~I. Bolozdynya, P.~L. Brink, R.~Bunker, B.~Cabrera, D.~O. Caldwell,
  J.~P. Castle, C.~L. Chang, R.~M. Clarke, M.~B. Crisler, P.~Cushman, A.~K.
  Davies, R.~Dixon, D.~D. Driscoll, L.~Duong, J.~Emes, R.~Ferril, R.~J.
  Gaitskell, S.~R. Golwala, M.~Haldeman, J.~Hellmig, M.~Hennessey, D.~Holmgren,
  M.~E. Huber, S.~Kamat, M.~Kurylowicz, A.~Lu, R.~Mahapatra, V.~Mandic, J.~M.
  Martinis, P.~Meunier, N.~Mirabolfathi, S.~W. Nam, H.~Nelson, R.~Nelson, R.~W.
  Ogburn, J.~Perales, T.~A. Perera, M.~C. Perillo~Isaac, W.~Rau, A.~Reisetter,
  R.~R. Ross, T.~Saab, B.~Sadoulet, J.~Sander, C.~Savage, R.~W. Schnee, D.~N.
  Seitz, T.~A. Shutt, G.~Smith, A.~L. Spadafora, J.-P.~F. Thompson, A.~Tomada,
  G.~Wang, S.~Yellin, and B.~A. Young, ``New results from the cryogenic dark
  matter search experiment,'' {\em Physical Review D} {\bfseries 68} no.~8,
  (2003) 082002.

\bibitem{ahmed010}
Z.~Ahmed, D.~Akerib, S.~Arrenberg, C.~Bailey, D.~Balakishiyeva, L.~Baudis,
  D.~Bauer, P.~Brink, T.~Bruch, R.~Bunker, B.~Cabrera, D.~Caldwell, J.~Cooley,
  P.~Cushman, M.~Daal, F.~DeJongh, M.~Dragowsky, L.~Duong, S.~Fallows,
  E.~Figueroa-Feliciano, J.~Filippini, M.~Fritts, S.~Golwala, D.~Grant,
  J.~Hall, R.~Hennings-Yeomans, S.~Hertel, D.~Holmgren, L.~Hsu, M.~Huber,
  O.~Kamaev, M.~Kiveni, M.~Kos, S.~Leman, R.~Mahapatra, V.~Mandic, K.~McCarthy,
  N.~Mirabolfathi, D.~Moore, H.~Nelson, R.~Ogburn, A.~Phipps, M.~Pyle, X.~Qiu,
  E.~Ramberg, W.~Rau, A.~Reisetter, T.~Saab, B.~Sadoulet, J.~Sander, R.~Schnee,
  D.~Seitz, B.~Serfass, K.~Sundqvist, M.~Tarka, P.~Wikus, S.~Yellin, J.~Yoo,
  B.~Young, and J.~Zhang, ``Dark matter search results from the cdms ii
  experiment,'' {\em Science} {\bfseries 327} no.~5973, (2010) 1619--1621.

\bibitem{bernabei03}
R.~Bernabei, P.~Belli, F.~Cappella, R.~Cerulli, F.~Montechia, F.~Nozzoli,
  A.~Incicchitti, D.~Prosperi, C.~J. Dai, H.~H. Kuang, J.~M. Ma, and Z.~P. Ye,
  ``Dark matter search,'' {\em Rivista Nuovo Cimento} {\bfseries 26} no.~1,
  (2003) 1--74.

\bibitem{bernabei08}
R.~Bernabei, P.~Belli, A.~Bussolotti, F.~Cappella, R.~Cerulli, C.~J. Dai,
  A.~D'Angelo, H.~L. He, A.~Incicchitti, H.~H. Kuang, J.~M. Ma, A.~Mattei,
  F.~Montecchia, F.~Nozzoli, D.~Prosperi, X.~D. Sheng, and Z.~P. Ye, ``The
  dama/libra apparatus,'' {\em Nuclear Instruments and Methods in Physics
  Research Section A} {\bfseries 592} no.~3, (2008) 297--315.

\bibitem{angle09}
J.~Angle, E.~Aprile, F.~Arneodo, L.~Baudis, A.~Bernstein, A.~Bolozdynya,
  L.~C.~C. Coelho, C.~E. Dahl, L.~Deviveiros, A.~D. Ferella, L.~M.~P.
  Fernandes, S.~Fiorucci, R.~J. Gaitskell, K.~L. Giboni, R.~Gomez, R.~Hasty,
  L.~Kastens, J.~Kwong, J.~A.~M. Lopes, N.~Madden, A.~Manalaysay, A.~Manzur,
  D.~N. McKinsey, M.~E. Monzani, K.~Ni, U.~Oberlack, J.~Orboeck, G.~Plante,
  R.~Santorelli, J.~M.~F. Dos~Santos, P.~Shagin, T.~Shutt, P.~Sorensen,
  S.~Schulte, C.~Winant, and M.~Yamashita, ``Constraints on inelastic dark
  matter from xenon10,'' {\em Physical Review D} {\bfseries 80} no.~11, (2009)
  115005.

\bibitem{summer05}
T.~Summer, ``The zeplin iii dark matter project,'' {\em New Astronomy Reviews}
  {\bfseries 49} no.~2-6, (2005) 277--281.

\bibitem{schnee011}
R.~W. Schnee, ``Introduction to dark matter experiments,'' in {\em Physics of
  the Large and Small: Proceedings of the 2009 Theoretical Advanced Study
  Institute in Elementary Particle Physics}, C.~Csaki and S.~Dodelson, eds.,
  pp.~775--830.
\newblock World Scientific, Singapore, 2011.

\bibitem{lee56}
T.~D. Lee and C.-N. Yang, ``Question of parity conservation in weak
  interactions,'' {\em Physical Review} {\bfseries 104} no.~1, (1956) 254--258.

\bibitem{kobzarev66}
I.~Kobzarev, L.~B. Okun$^{'}$, and I.~Pomeranchuk, ``On the possibility of
  observing mirror particles,'' {\em Soviet Journal of Nuclear Physics}
  {\bfseries 3} (1966) 837.

\bibitem{pavsic74}
M.~Pav\v{s}i\v{c}, ``External inversion, internal inversion, and reflection
  invariance,'' {\em International Journal of Theoretical Physics} {\bfseries
  9} no.~4, (1974) 229--244.

\bibitem{blinnikov83}
S.~I. Blinnikov and M.~Y. Khlopov, ``Possible astronomical effects of mirror
  particles,'' {\em Soviet Astronomy} {\bfseries 27} no.~4, (1983) 371--375.

\bibitem{foot91}
R.~Foot, H.~Lew, and R.~R. Volkas, ``A model with fundamental improper
  spacetime symmetries,'' {\em Physics Letters B} {\bfseries 272} no.~1-2,
  (1991) 67--70.

\bibitem{khlopov91}
M.~Y. Khlopov, G.~M. Beslin, N.~G. Bochkarev, L.~A. Pustil$^{'}$nik, and S.~A.
  Pustil$^{'}$nik, ``Observational physics of the mirror world,'' {\em Soviet
  Astronomy} {\bfseries 35} no.~1, (1991) 21--29.

\bibitem{khlopov99}
M.~Y. Khlopov, {\em Cosmoparticle physics}.
\newblock World Scientific, Singapore, 1999.

\bibitem{silagadze01}
Z.~K. Silagadze, ``Tev scale gravity, mirror universe and$\ldots$dinosaurs,''
  {\em Acta Physica Polonica B} {\bfseries 32} no.~1, (2001) 99--128.

\bibitem{foot04}
R.~Foot, ``Mirror matter-type dark matter,'' {\em International Journal of
  Modern Physics D} {\bfseries 13} no.~10, (2004) 2161--2192.

\bibitem{foot06}
R.~Foot, ``Generalized mirror matter models,'' {\em Physics Letters B}
  {\bfseries 632} no.~4, (2006) 467--470.

\bibitem{okun07}
L.~B. Okun$^{'}$, ``Mirror particles and mirror matter: 50 years of speculation
  and searching,'' {\em Physics-Uspekhi} {\bfseries 50} no.~4, (2007) 380--389.

\bibitem{ciarcelluti010}
P.~Ciarcelluti, ``Cosmology with mirror dark matter,'' {\em International
  Journal of Modern Physics D} {\bfseries 19} no.~14, (2010) 2151--2230.

\bibitem{foot02}
R.~Foot, {\em Shadowlands: Quest For Mirror Matter In The Universe}.
\newblock Universal Publishers, Parkland, 2002.

\bibitem{ignatiev00}
A.~Y. Ignatiev and R.~R. Volkas, ``Geophysical constraints on mirror matter
  within the earth,'' {\em Physical Review D} {\bfseries 62} no.~2, (2000)
  023508.

\bibitem{foot01a}
R.~Foot, ``The mirror world interpretation of the 1908 tunguska event and other
  more recent events,'' {\em Acta Physica Polonica B} {\bfseries 32} no.~10,
  (2001) 3133--3146.

\bibitem{foot01}
R.~Foot and Z.~K. Silagadze, ``Do mirror planets exist in our solar system?,''
  {\em Acta Physica Polonica B} {\bfseries 32} no.~7-8, (2001) 2271--2278.

\bibitem{silagadze02}
Z.~K. Silagadze, ``Mirror objects in the solar system?,'' {\em Acta Physica
  Polonica B} {\bfseries 33} no.~5, (2002) 1325--1341.

\bibitem{silagadze05}
Z.~K. Silagadze, ``Tunguska genetic anomaly and electrophonic meteors,'' {\em
  Acta Physica Polonica B} {\bfseries 36} no.~3, (2005) 935--964.

\bibitem{foot03a}
R.~Foot and S.~Mitra, ``Mirror matter in the solar system: new evidence for
  mirror matter from eros,'' {\em Astroparticle Physics} {\bfseries 19} no.~6,
  (2003) 739--753.

\bibitem{foot03b}
R.~Foot and S.~Mitra, ``Have mirror micrometeorites been detected?,'' {\em
  Physical Review D} {\bfseries 68} no.~7, (2003) 071901.

\bibitem{dietl011}
C.~Dietl, L.~Labun, and J.~Rafelski, ``Properties of gravitationally bound dark
  compact ultra dense objects,'' {\em Physics Letters B} (2012) ,
  \href{http://arxiv.org/abs/1110.0551}{{\ttfamily arXiv:1110.0551
  [astro-ph.CO]}}.

\bibitem{labun011}
L.~Labun, J.~Birrell, and J.~Rafelski, ``Solar system signatures of impacts by
  compact ultra dense objects,''
  \href{http://arxiv.org/abs/1104.4572}{{\ttfamily arXiv:1104.4572
  [astro-ph.EP]}}.

\bibitem{coleman85}
S.~Coleman, ``Q-balls,'' {\em Nuclear Physics B} {\bfseries 262} no.~2, (1985)
  263--283.

\bibitem{kusenko97}
A.~Kusenko, ``Solitons in the supersymmetric extensions of the standard
  model,'' {\em Physics Letters B} {\bfseries 405} no.~1-2, (1997) 108--113.

\bibitem{kusenko98}
A.~Kusenko, ``Q-balls in the mssm,'' {\em Nuclear Physics B Proceedings
  Supplements} {\bfseries 62} no.~1-3, (1998) 248--252.

\bibitem{zhitnitsky06}
A.~Zhitnitsky, ``Cold dark matter as compact composite objects,'' {\em Physical
  Review D} {\bfseries 74} no.~4, (2006) 043515.

\bibitem{cumberbatch08}
D.~T. Cumberbatch, J.~Silk, and G.~D. Starkman, ``Difficulties in explaining
  the cosmic photon excess with compact composite object dark matter,'' {\em
  Physical Review D} {\bfseries 77} no.~6, (2008) 063522.

\bibitem{kusenko09}
A.~Kusenko and I.~M. Shoemaker, ``Neutrinos from the terrestrial passage of
  supersymmetric dark-matter q-balls,'' {\em Physical Review D} {\bfseries 80}
  no.~2, (2009) 027701.

\bibitem{will93}
C.~Will, {\em Theory and experiment in gravitational physics. Revised edition}.
\newblock Cambridge University Press, Cambridge, 1993.

\bibitem{roy05}
A.~E. Roy, {\em Orbital Motion}.
\newblock Institute of Physics, Bristol, fourth edition~ed., 2005.

\bibitem{gillessen09}
S.~Gillessen, F.~Eisenhauer, T.~K. Fritz, H.~Bartko, K.~Dodds-Eden, O.~Pfuhl,
  T.~Ott, and R.~Genzel, ``The orbit of the star s2 around sgr a$^{\ast}$ from
  very large telescope and keck data,'' {\em The Astrophysical Journal}
  {\bfseries 707} no.~2, (2009) L114--L117.

\bibitem{gillessen012}
S.~Gillessen, R.~Genzel, T.~K. Fritz, E.~Quataert, C.~Alig, A.~Burkert,
  J.~Cuadra, F.~Eisenhauer, O.~Pfuhl, K.~Dodds-Eden, C.~F. Gammie, and T.~Ott,
  ``A gas cloud on its way towards the supermassive black hole at the galactic
  centre,'' {\em Nature} {\bfseries 481} no.~7379, (2012) 51--54.

\bibitem{li08}
G.~Li, Z.~Yi, G.~Heinzel, A.~R\"{u}diger, O.~Jennrich, L.~Wang, Y.~Xia,
  F.~Zeng, and H.~Zhao, ``Methods for orbit optimization for the lisa
  gravitational wave observatory,'' {\em International Journal of Modern
  Physics D} {\bfseries 17} no.~7, (2008) 1021--1042.

\bibitem{lass83}
H.~Lass and L.~Blitzer, ``The gravitational potential due to uniform disks and
  rings,'' {\em Celestial Mechanics and Dynamical Astronomy} {\bfseries 30}
  no.~3, (1983) 225--228.

\bibitem{roth72}
E.~A. Roth, ``Perturbation of a test body by a massive ring in a newtonian
  field,'' Special Report~21, European Space Research Organisation, 1972.

\bibitem{kasatkin05}
G.~V. Kasatkin, ``Internal gravitational field of a thin homogeneous ring,''
  {\em Cosmic Research} {\bfseries 43} no.~4, (2005) 245--253.

\bibitem{Lask010}
J.~Laskar and G.~Bou\'{e}, ``Explicit expansion of the three-body disturbing
  function for arbitrary eccentricities and inclinations,'' {\em Astronomy and
  Astrophysics} {\bfseries 522} (November, 2010) A60.

\bibitem{lense18}
J.~Lense and H.~Thirring, ``\"{U}ber den einflu{\ss} der eigenrotation der
  zentralk\"{o}rper auf die bewegung der planeten und monde nach der
  einsteinschen gravitationstheorie,'' {\em Physikalische Zeitschrift}
  {\bfseries 19} (1918) 156--163.

\bibitem{pijpers98}
F.~P. Pijpers, ``Helioseismic determination of the solar gravitational
  quadrupole moment,'' {\em Monthly Notices of the Royal Astronomical Society}
  {\bfseries 297} no.~3, (1998) L76--L80.

\bibitem{Roz011}
J.-P. Rozelot and C.~Damiani, ``History of solar oblateness measurements and
  interpretation,'' {\em The European Physical Journal H} {\bfseries 36} no.~3,
  (2011) 407--436.

\bibitem{williams84}
J.~Williams, ``Determining asteroid masses from perturbations on mars,'' {\em
  Icarus} {\bfseries 57} no.~1, (1984) 1--13.

\bibitem{standish02}
E.~M. Standish and A.~Fienga, ``Accuracy limit of modern ephemerides imposed by
  the uncertainties in asteroid masses,'' {\em Astronomy and Astrophysics}
  {\bfseries 384} no.~1, (2002) 322--328.

\bibitem{fienga05}
A.~Fienga and J.-L. Simon, ``Analytical and numerical studies of asteroid
  perturbations on solar system planet dynamics,'' {\em Astronomy and
  Astrophysics} {\bfseries 429} no.~1, (2005) 361--367.

\bibitem{mouret09}
S.~Mouret, J.~L. Simon, F.~Mignard, and D.~Hestroffer, ``The list of asteroids
  perturbing the mars orbit to be seen during future space missions,'' {\em
  Astronomy and Astrophysics} {\bfseries 508} no.~1, (2009) 479--489.

\bibitem{souchay09}
J.~Souchay, D.~Gauchez, and A.~Nedelcu, ``Influence of the largest asteroids on
  the orbital motions of terrestrial planets: application to the earth and
  mars,'' {\em Solar System Research} {\bfseries 43} no.~1, (2009) 79--81.

\bibitem{somenzi010}
L.~Somenzi, A.~Fienga, J.~Laskar, and P.~Kuchynka, ``Determination of asteroid
  masses from their close encounters with mars,'' {\em Planetary and Space
  Science} {\bfseries 58} no.~5, (2010) 858--863.

\bibitem{su99}
Z.-Y. Su, A.-M. Wu, D.~Lee, W.-T. Ni, and S.-C. Lin, ``Asteroid perturbations
  and the possible determination of asteroid masses through the astrod space
  mission,'' {\em Planetary and Space Science} {\bfseries 47} no.~3-4, (1999)
  339--343.

\bibitem{vinet06}
J.-Y. Vinet, ``Lisa and asteroids,'' {\em Classical and Quantum Gravity}
  {\bfseries 23} no.~15, (2006) 4939--4944.

\bibitem{mouret011}
S.~Mouret, ``Tests of fundamental physics with the gaia mission through the
  dynamics of minor planets,'' {\em Physical Review D} {\bfseries 84} no.~12,
  (2011) 122001.

\bibitem{carry08}
B.~Carry, C.~Dumas, M.~Fulchignoni, W.~J. Merline, J.~Berthier, D.~Hestroffer,
  T.~Fusco, and P.~Tamblyn, ``Near-infrared mapping and physical properties of
  the dwarf-planet ceres,'' {\em Astronomy and Astrophysics} {\bfseries 478}
  no.~1, (2008) 235--244.

\bibitem{pitjeva05b}
E.~V. Pitjeva, ``Relativistic effects and solar oblateness from radar
  observations of planets and spacecraft,'' {\em Astronomy Letters} {\bfseries
  31} no.~5, (2005) 340--349.

\end{thebibliography}\endgroup
%-----------------------------------------

\end{document}